\documentclass[superscriptaddress,amsmath,twocolumn,amssymb,aps,showpacs,prb]{revtex4-1}
\usepackage[utf8]{inputenc}
\usepackage{graphicx}
\usepackage{bm}
\usepackage{amssymb}
\usepackage{amsfonts}
\usepackage{color}
\usepackage{natbib}
\usepackage[caption=false]{subfig}
\newcommand{\vep}{\varepsilon}

\newcommand{\bqa}{\begin{eqnarray}}
\newcommand{\eqa}{\end{eqnarray}}
\newcommand{\nn}{\nonumber}

\usepackage{atbegshi,picture}
\usepackage{lipsum}

\AtBeginShipoutNext{\AtBeginShipoutUpperLeft{%
		\put(\dimexpr\paperwidth-.3cm\relax,-1.cm){\makebox[0pt][r]{{
		 Ann. Phys. 456C, 169306 (2023), Special Issue of Annals of Physics  on Correlation and Disorder in Memory of Konstantin B. Efetov
}}}%
	}}

\begin{document}

\date{\today}

\title{Towards a Comprehensive  Theory  of    Metal-Insulator Transitions
in Doped Semiconductors}

\author{S. Kettemann}
\affiliation{ Constructor University (former Jacobs University)
  Bremen, Bremen 28759, Germany}

%
%

\begin{abstract}
A review is given on   the  theory of  metal-insulator transitions  (MIT) 
in doped semiconductors. We  
focus in particular on reviewing   theories  of    their anomalous magnetic properties, 
which emerge from  the interplay of  spin and charge   correlations and disorder.
   Building on the review of these  existing theories and experiments,  we  suggest  that the 
    finite temperature phase diagram can  be structured into       
   1. a quantum spin liquid phase 
   at subcritical doping and low temperature, 
   as described  by the Bhatt-Lee theory of random  spin clusters,
    mostly random singlets. 2.
   a 
   critical non-Fermi-liquid fan, originating at the MIT, 
     which is dominated by random Kondo singlets with a universal tail of the distribution of
      their binding energies.  This is caused by   multifractality and 
      results  in an anomalous power law divergence 
      of the magnetic susceptibility with a universal power 
     and 3.   a supercritical, low temperature  phase.  
        Rare events caused by the random placement of dopants 
        do not allow to define strict phase boundaries.   
        Remaining  open problems are reveiwed and outlined.  
Finally, the possibility of   finite temperature 
 delocalization transitions is reviewed, which are 
  caused by the correlation induced temperature dependence of the spin scattering rate from 
 magnetic moments. 
This review article is devoted to the memory of  Konstantin B. Efetov. 
\end{abstract}

\maketitle

\section{Introduction}

Doped semiconductors, such as 
 ${\rm Si_{1-x}P_x} $ (Si:P),   are well known to show a 
  metal-insulator transition  (MIT) by variation of the  dopant concentration $x$. 
Although 
  silicon is one of the  most intensively studied materials in human history
  and   the  MIT  in
 doped silicon 
  has been long known as a paradigm of 
 quantum phase transitions,
    it remains only partially understood 
      \cite{Rosenbaum83,Milligan1985,lohneysen2011,localisation2020}.
While  the doping  increases  the charge carrier density and thereby the  overlap integral between 
dopant wave functions,
 the random positioning   of   the dopants
results  in random hopping amplitudes between  dopant sites.
 Randomness  causes the  charge carriers
 to  remain     localized   below a critical dopant density  \cite{pwanderson}, the celebrated Anderson localization. 
   However,  there  are indications of  strong  spin and charge correlations at 
    low dopant densities and  in the vicinity of the MIT, where  the 
    electron-electron interaction is poorly screened, giving rise to anomalous 
    magnetic, transport and  optical properties.  
   Therefore, while 
  the  transition to a metal in doped semiconductors 
    is found to be very similar to an  Anderson MIT (AMIT) \cite{pwanderson}, driven  by  disorder, rather than a 
 correlation-driven Mott transition \cite{mott},
    there is  experimental evidence for  non-Fermi liquid 
    behavior due to the presence of local magnetic  moments (MMs). 
    Thus,  a theory of  the MIT  in doped semiconductors as function of dopant concentration or pressure
    requires to fully model both disorder \cite{pwanderson,wegner76,Lee1985,kramermackinnon,Belitz1994,Efetov1983,Efetov1997,Evers2008}, 
    and spin and charge correlations \cite{Andres1981,bhattlee81,bhattlee82,Lee1985,Paalanen1988,Sachdev1989,Belitz1994,mott,BhattFisher92,Lakner1994,Miranda05,lohneysen2011}.

         This makes 
    the derivation of its critical properties
  one of the most  demanding   challenges in condensed matter theory \cite{Lee1985,kramermackinnon,Belitz1994,Evers2008}. 
  Getting a more complete understanding of the MIT in 
   one of the best studied materials, silicon, will 
improve the understanding of the MIT in other, more complex materials, 
 such as SrTiO3 \cite{lin}  or the cuprates \cite{Norman2003,cuprates,Lee2006,Fradkin2015,Pepin2023} and will 
 be of crucial relevance
  for understanding the doping dependence in  topological insulators \cite{hasankane}, 
    topological superconductors \cite{foster}, and the 
   superconductor-insulator transition 
    in doped semiconductors \cite{lin,strunk}. 
    Moreover a better understanding of the interplay of disorder and spin and charge  correlations 
    in doped semiconductors may improve their design as  functional materials, such as 
    dilute magnetic semiconductors
      \cite{Sato2010}, high efficiency solar cell concepts, like  intermediate band solar cells   \cite{Nice1995,Marti1997,Okada2015} or  dye sensitized solar cells
 \cite{Nelson1999} and   thermoelectric devices 
 \cite{Agne2021,Castellani1988,Villagonzalo1999}.

\section{
Scaling Theory }
 The MIT in doped semiconductors  is  consistent with   a second order quantum phase transition where the  localization length  and correlation length 
   diverge,  on the respective sides of the transition \cite{stupp,bogdanovich,itoh}, even though the possibility of a discontinuous transition with a minimum metallic conductivity  \cite{Mott1961}  is also considered  \cite{Moebius}. Measurements of conductivity are found to 
   be consistent with  Wegner  scaling \cite{wegner76}
 with  dopant density $N$   at finite temperature $T$,
 $ \sigma(N,T) = \xi (N)^{2-d} f  [T/\Delta_{\xi} ],$  where  $f[x]$ is a scaling function. This  yields with the correlation length 
 $\xi(N)  \sim |N-N_c|^{-\nu}$  with the critical dopant density $N_c$ and the corresponding energy scale $\Delta_{\xi} \sim \xi(N)^{-z}$,where $z$ is the dynamical exponent, for  dimension $d=3$
      \begin{equation} \label{scaling}
       \sigma(N,T) =
     \left( \frac{|N-N_c|}{N_c} \right)^{\mu}  F [ T  \left( \frac{|N-N_c|}{N_c} \right)^{-z \nu}],
 \end{equation}
where $F[x]$ is a  scaling function.  
 It follows that  at criticality the conductivity scales with temperature as  $  \sigma(N_c,T) = \sigma_c(T)  \sim T^{\mu/(z \nu)}.$ 
 Wegner scaling corresponds to $\mu = \nu,$
 but in fitting the experiments $\mu$ is often allowed to be an independent parameter \cite{bogdanovich}.
  For the Anderson transition of noninteracting disordered electrons, $\Delta_{\xi}$ is on the insulating side of the transition the local level spacing, so that $z=d$ and $\mu= \nu$.  It should be noted hat the one parameter scaling  of the conductivity Eq. (\ref{scaling})
  is only justified, if 
  the localisation length is self averaging, 
 that is, if it  has a normal distribution, which has been confirmed
  for the Anderson localisation transition numerically, for a review see Ref.  \cite{kramermackinnon}.
  
The scaling of conductivity of uncompensated bulk ($d=3$)
 Si:P was done in  Ref.  \cite{stupp}, where the authors assumed  $\mu = \nu$ and found 
 $z \approx 2.4$ and $\nu  \approx 1.3$. 
  In Ref.  \cite{bogdanovich} the MIT in Si:B was studied as function of stress. They find $\mu  \approx 1.6$ and $z \nu  \approx 3.2$. Assuming $\mu = \nu$,
   this yields thus $\nu  \approx 1.6$ and $z \approx 2$. 
  In Ref.   \cite{itoh} the MIT in p-doped uncopmensated  semiconductor,  Ge:Ga has been 
  studied as function of doping concentration. They find $\mu = \nu \approx 1.2$ for $z\approx 3$. 
  For compensated Ge:Ga,As,  they rather find  $\mu = \nu \approx 1$ for $z \approx 3.$
    Earlier measurements  gave  $\nu \approx 1$ \cite{paalanenthomas,kramermackinnon}.
  
       The  scaling theory of the 
       Anderson transition was 
       conjectured in Ref.  \cite{Abrahams79} 
     in terms of  the flow of the dimensionless conductance $g$ with  system size $L$,
   building on Refs.  \cite{Thouless74,wegner76}.
Wegner formulated that scaling theory in   a field-theoretical description 
 in terms of the nonlinear  sigma model \cite{Wegner79}.  This  effective model for  the  long wave length physics, which captures in particular diffusion modes originating from 
 multiple impurity scattering 
   was then 
  derived for disordered electron systems, 
  performing  the disorder averaging nonperturbatively with   the Replica trick  in Refs.   \cite{Efetov1980,Juengling1980,Schaefer1980}
and   with  the 
supersymmetry method  by Efetov  \cite{Efetov1983}, both without and with magnetic field and in the presence of spin-orbit coupling and magnetic impurities. 
A  resummation of singularities in perturbation theory 
 provided further evidence of Anderson localisation in 
  $d\leq 2$ dimensions without magnetic field \cite{Gorkov1979},
  \cite{Vollhardt1980}.
  The Anderson transition was studied  in $d= 2 + \epsilon$ expansion  of the nonlinear sigma model upto 4-loop order \cite{Hikami1983}, \cite{Wegner1989} and 5-loop order \cite{Hikami1992}, 
  on the Bethe lattice  \cite{Efetov1985,Efetov1987, Zirnbauer1986,Mirlin1991} and in effective medium approximation
   \cite{Efetov1990} and  \cite{Fyodorov1992}, yielding in $d=3$ estimates for the critical exponent $\nu \approx 1$. Recently the conformal bootstrap method has been applied to the Anderson transition in 2D with spin-orbit interaction, and suggested to be a way to get more accurate estimates for the critical exponent $\nu$ also for the 3D Anderson transition \cite{Hikami2018}.
   
Numerical transfer matrix calculations 
        for the 3D  Anderson  tight binding (Atb) model of noninteracting 
         disordered fermions  \cite{kramermackinnon}  gave for the critical exponent 
       without magnetic field
    $\nu = 1.571 \pm .004$ \cite{slevin2014}
    and        finite size scaling of the
  distribution of wave function intensities yields $\nu = 1.59 \pm .006$   \cite{rodriguez2011}.  
  
    The discrepancy with the  experimental results  could be  due  to
    interaction effects. 
     Local interactions can create magnetic moments in weakly coupled impurity sites   \cite{andersonmm,Toyozawa62,bhatt88}. 
     In Ref.  \cite{Jung2016}  the 3D  Atb model of noninteracting 
         disordered fermions with  hopping amplitude $t$  
  was studied  when the electron spin is  coupled by exchange coupling $J$ to 
  a finite concentration of classical magnetic moments. 
    The critical exponent was found for  $J> 0.3 t,$      to be  $\nu_S \approx 1.3 \pm 0.1$ 
     when  5 percent of all lattice sites  have magnetic moments  \cite{Jung2016}, 
     which is in fair agreement with experiments on uncompensated bulk ($d=3$)
 Si:P\cite{stupp}.
 
  A density functional theory analysis of  a model 
      of randomly distributed  donor impurities with long range Coulomb  interactions between 
      spinless 
       charge carriers yields 
        $\nu = 
       1.30(+0.12, - 0.06)$ 
    \cite{harashima2014},which is also  in  good 
    agreement with    experimental results \cite{stupp}, as well. 
    
   A self consistent  Hartree Fock treatment of 
    disordered electrons with long range interactions yields a
     smaller $\nu,$ which is found to depend
      on the interaction range \cite{amini} and 
   for the diverging  correlation length   a different $\nu_M$ was obtained 
   than for the diverging localization length $\nu_I$ on the 
     insulator 
    side of the MIT.
   
    A
     typical medium dynamical cluster approximation for disordered electronic systems
      has been applied to  the disordered Anderson-Hubbard model, a model 
       with onsite interactions,
     identifying  the mobility edge and  the  finite temperature phase diagram 
     \cite{Terletska2018}. 
     	Parameter-free ab-initio calculations of doped Si:P
	which employed density functional theory to build a tight binding model, 
	that is then diagonalised numerically,  
	 have been used to address the 
	exponent puzzle for the Anderson transition in 
	both compensated and uncompensated 
	doped semiconductors in Ref. 
  \cite{Carnio2019a}.
      
    Cold atom experiments on the  Anderson localization  transition in $d=3$
       reported critical exponents
 $\nu \approx 1.6$
 \cite{Chabe2008QKM,lopez12}. In these systems  interactions are known to be weak,  
 which explains the  good agreement with the results of the noninteracting tight binding model \cite{slevin2014,rodriguez2011}.

     The many-body 
  theory of the MIT in doped semiconductors has a 
   long history.
It is   well established  \cite{Lee1985,Belitz1994} that   long range Coulomb interactions  in strongly disordered metals 
  give rise to  Althshuler-Aronov corrections to the  density of states (DOS)
  and to the temperature dependence of the conductivity on the metallic side of the MIT
    In 3D it  gives 
     $ \sigma \sim \sigma_0 - \sigma_0^2 A T^{1/2},$ 
     where $\sigma_0$
     is the residual conductivity and $A$  an interaction dependent constant.
      This corresponds
    according to  Eq. (\ref{scaling})  for $\mu = \nu$ 
    to a dynamical exponent $z=2$ \cite{altshuleraronov,finkelstein,finkelstein88}.
    For a review see Ref.  \cite{Schwiete2023}.
    This is in strong contrast to the Fermi liquid 
     conductivity of a   pure metal, where the electron-electron scattering rate
     is proportional to $T^2$, which would rather yield  $z=1/2$. 
  On the insulator side of the MIT  Coulomb interactions give rise to  
 the 
 Coulomb gap in  the DOS.  Transport of electrons 
 can be modeled there by   Efros-Shklovskii  variable range hopping 
 conductivity \cite{efros}, which  is consistend with Wegner scaling for $N < N_c$,
   when $\sigma (N,T) =  \sigma_c(T) \exp ( -(\Delta_{\xi} (N)/T)^{p})$,
   where $\Delta_{\xi} (N) \sim (N_c-N)^{z' \nu}$. Here, the dynamical exponent $z'$
    should, for consistency, be  equal to $z$. In fitting experiments it is sometimes taken as  an independent fitting parameter. Here, $p=1/2$ or $p=1/4$ depending on the dominant hopping process \cite{efros}.
      \begin{figure}[t]
\includegraphics[width=8cm,angle=0]{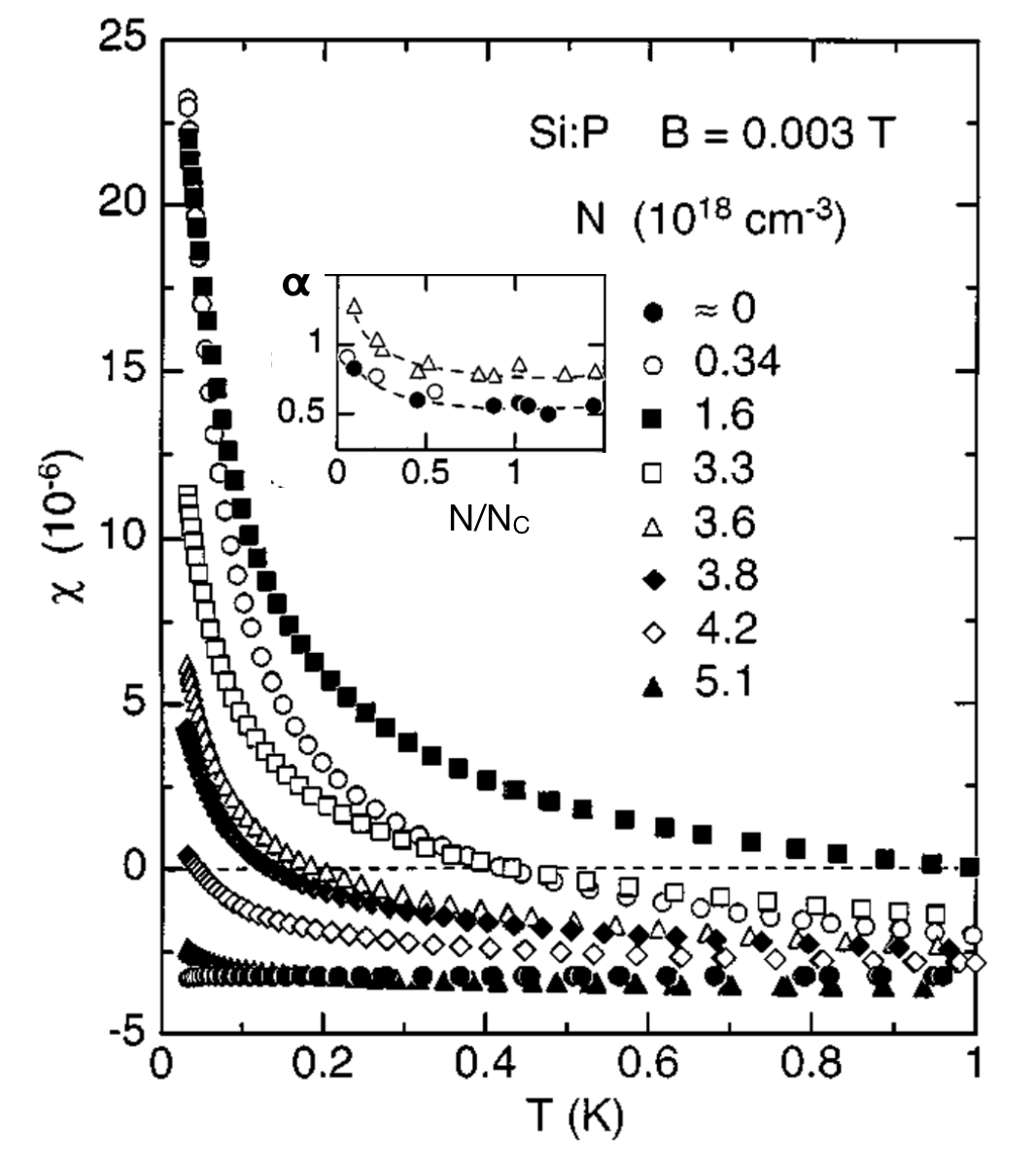}
\caption{  
 Measured magnetic susceptibility    for various doping concentrations
 $N$ \cite{Schlager1997}. Inset: Power $\alpha,$ as obtained by fitting    $\chi(T) \sim T^{-\alpha}$  (circles), and by fitting the   magnetic moment contribution to  the specific heat to   $C(T) \sim T^{1-\alpha}$  (triangles), 
  plotted as function of $N/N_c$.  
    Figs. taken and reedited from 
 Ref.   \cite{Schlager1997},  Copyright  1997 Institute of Physics Science.
        }
\label{suscsip}
\end{figure}
  
 \section{Anomalous Magnetic Properties}
 \subsection{Experiments}
Magnetic susceptibility and specific heat measurements of 
 doped 
 semiconductors like Si:P  \cite{Quirt,Kobayashi,Paalanen1988,Lakner1994,Schlager1997,Alloul,Roy,lohneysen2011}
  as well as ESR measurements \cite{Hirsch}
  provide  evidence for magnetic moments  (MMs).
 The magnetic susceptibility 
 is experimentally found to diverge  at 
   low temperature with a power law
      $ \chi \sim T^{-\alpha}$, 
     see  Fig. \ref{suscsip}.    The fitted
  power $\alpha$ is plotted as function of 
          dopant density $N$ in the inset of Fig. \ref{suscsip} (circles), 
           while an analysis of the contribution of MMs to the  specific heat measurements yields
           $\alpha(N)$  in the inset of  Fig. \ref{suscsip} (triangles)   \cite{Schlager1997,lohneysen2011}.  
    In the dilute limit $N \le 1/a_B^3$,
     where $a_B$ is the Bohr radius of the dopants,
     the magnetic susceptibility is observed to follow the Curie law
$\chi \sim N/T.$ This is in agreement with the fact that   in Si:P
     all dopants have a free $S=1/2$-spin,  only 
    weakly coupled by the antiferromagnetic super-exchange interaction
     between the hydrogen like dopant levels   \cite{Andres1981}.
     For increasing doping concentration $N$, the  power $\alpha$ is found to decrease
     continously. 
         Close to   the transition 
   $ N_c =  3.3 \times 10^{18}/{\rm cm^3},$
              the power converges to    $\alpha \approx .5,$
        and   hardly depends on   $N$ deep into the metallic phase.
        At the same time,  
         the amplitude of the  magnetic susceptibility 
           decays with increasing 
            concentration $N,$  indicating a decrease of the concentration of MMs $n_M< N$. 

At the MIT $N= N_c$ and  in the metal regime  $N> N_c$
 a finite concentration of  spin$-1/2$ MMs have been shown to be   created 
  by the  onsite interaction in weakly coupled impurity sites   \cite{andersonmm,Toyozawa62,bhatt88}, 
  positioned randomly and  
     coupled weakly by  antiferromagnetic super-exchange interactions \cite{mott,bhattlee82,lohneysen2011}.   
           The concentration of MMs decreases with increasing   doping. 
    At the MIT theoretical calculations 
     \cite{Toyozawa62,Milovanovic1989}
    and  the evaluation of experiments     \cite{lohneysen2011}  are consistent with
     about 10 \% of all dopants forming MMs. In Ref.       \cite{finkelsteinesr} it was pointed out that 
     a combined analysis of magnetic susceptibility and electron spin resonance experiments 
     gives hints that at least part of the low temperature enhancement of the magnetic susceptibility 
     could be  rather due to an anomalous intensification of the interaction between electrons by disorder. 
              
  In a metal  the exchange coupling $J$ of  spin $1/2$-MMs 
   to the conduction electrons results below a temperature scale $T_{\rm K}$
   in the screening of the MM
    by a cloud of conduction electron spins, the Kondo effect  \cite{Kondo64,Nagaoka65,Suhl65,Wilson75,bethe}.
Above the Kondo temperature $T_{\rm K}$ the MMs contributes a Curie like
term to the magnetic susceptibility.
At temperatures below $T_{\rm K}$ the contribution to the susceptibility 
converges to  a temperature independent Pauli like contribution since   the MMs become screened.
Nagaoka \cite{Nagaoka65}  and Suhl \cite{Suhl65} derived for a
non-disordered metal with a  density of states 
    $\rho$ at the Fermi energy $\vep_F$
the Kondo temperature $T_{\rm K}$ as
$T_{\rm K}^0 \approx  c ~\vep_F  \exp \left( - 1/(\rho J) \right),$ where $c\approx 1.14.$
  The temperature 
         dependence of 
     thermodynamic
        observables  like the magnetic susceptibility
         and transport properties
          obey  in a clean metal universal scaling functions, which scale with $T_K$.                            
  
Indeed, the Kondo effect  has been detected experimentally  in doped semiconductors 
   for $N>N_c$ by thermopower measurements which, when fitting with  the theory of Ref. 
    \cite{maki}, is consistent 
   with a Kondo temperature 
   $T_{\rm K} \approx 0.8 {\rm K}$  \cite{lakner}.  
 Therefore, it is experimentally evident that  the Kondo effect has to be taken into account in the theory of the anomalous magnetic properties of doped semiconductors in the vicinity of the MIT. 
 
While the Kondo problem has been previously solved in clean metals   
  \cite{Kondo64,Nagaoka65,Suhl65,Wilson75}, 
 a comprehensive theory of the MIT  requires to 
consider  
the Kondo impurities coupled to the   strongly disordered electron systems 
 in the vicinity of the MIT in doped semiconductors. 
 
\subsection{Distribution of the Kondo temperature} Since the Kondo temperature in a clean metal   \cite{Kondo64,Nagaoka65,Suhl65}
 depends on the product of the local  exchange coupling $J$ and the 
density of states  at the Fermi energy
 $\rho$, it is natural  to expect a distribution of the Kondo temperature, $P(T_K)$, when 
  $J$ and $\rho$ are distributed due to the random placement of the dopants. 
           Early theories derived  thereby $P(T_K)$ from the distribution of $J$, 
           assuming that the dopants are on random, uncorrelated lattice sites
              \cite{BhattFisher92,langenfeld}, finding
              a  magnetic susceptibility with  corrections to the Curie law, 
              $\chi(T) \sim \exp(-k N a^3 \ln^3 (T_0/T))/T,$ where $k$ is a constant,  $T_0 \sim Z  t,$
              where $Z$ is the average degree of the random network  of dopants, and 
               $t$ the average hybridisation energy between neighboured dopant sites
               (for Si:P $t\approx 28 {\rm meV}$ was estimated in Ref.  \cite{langenfeld}). 
              Inserting the 
               local density of states
               $\rho(\epsilon,x)$  which is   known  to have a wide, lognormal  distribution
               at the AMIT  into 
               $T_{\rm K} \sim \exp \left( - 1/(\rho J) \right)$,   $P(T_K)$, was derived in Ref.  \cite{dobros}, and shown to  yield also weak
          corrections to the Curie law    for the magnetic susceptibility \cite{dobros},
               \cite{Miranda05}.
                    \begin{figure}[h]
\includegraphics[width=0.38\textwidth]{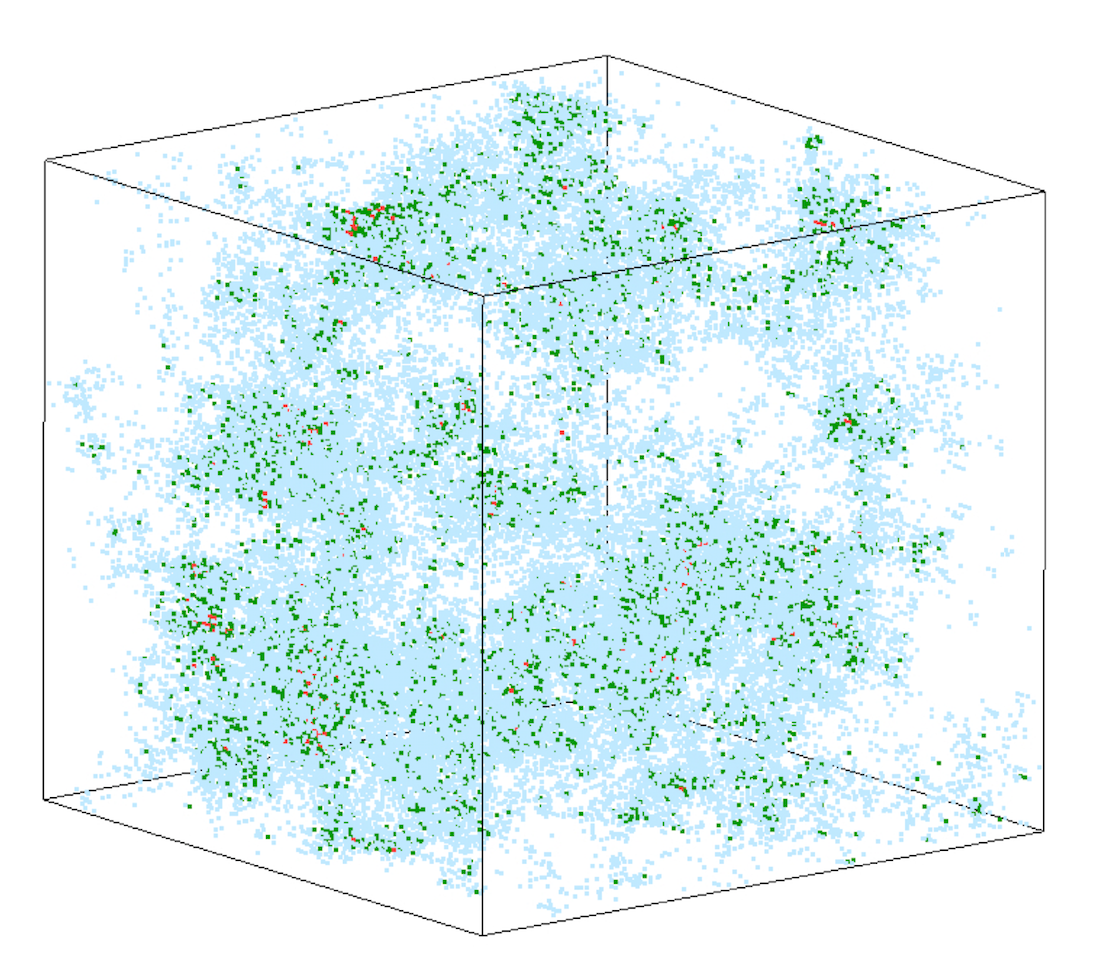}
\caption{
 Intensity of critical state of the 3D Atb model 
   (disorder amplitude 
    $W=16.5 t,$ $t$ hopping parameter, 
    at energy $E=2t$,
   size $L=100 a$, lattice spacing $a$).  Coloring denotes  $ \alpha_{\psi} = -\ln |\psi|^2 / \ln L \in [ 1.2,1.8], [1.8,2.4], [2.4,3.0]$, (red, green, light blue), respecitively.
    Fig. taken  from 
 Ref.   \cite{Kats},  Copyright  2012, American Physical Society.}
\label{mf}
\end{figure}

   \begin{figure}
\begin{center}
\includegraphics[width=0.45\textwidth]{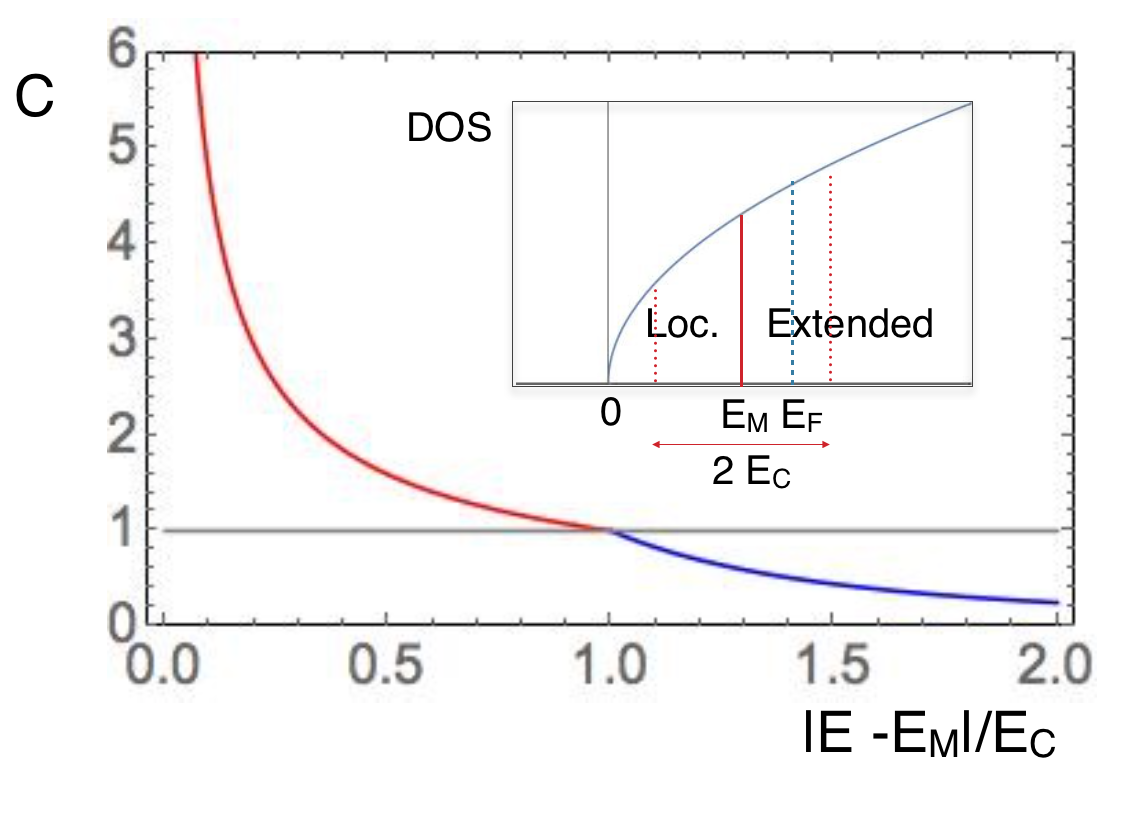}
\caption{  Correlation function of intensities, $C,$ 
as function of  energy difference $|E-E_M|$ in 
3D disordered system, $\eta \approx 2$. Inset: Density of states
   as function of  $E$. 
     Intensities are correlated within  energy  $ E_c$  around mobility edge $E_M$.
     Fig. taken  from 
 Ref.   \cite{PRL2016},  Copyright  2016, American Physical Society.}
\label{mfc}
\end{center}
\vspace{-25pt}
\end{figure}

           The influence of the disorder in the electron  system
           on the Kondo effect is more subtle, however. 
   In 1-loop approximation, 
   the Kondo
temperature at site ${\bf r}$ of a spin-$1/2$-impurity is  determined  by \cite{Nagaoka65,Suhl65}
\begin{equation}
\label{eq:FTK}
1= \frac{J}{2 N_L} \sum_{n=1}^{N_L} \frac{L^d |\psi_n ({\bf r})|^2}{E_n-E_F} \tanh \left(  
\frac{E_n-E_F}{2 T_{\rm K}}({\bf r}) \right),
\end{equation}
where $N_L$ is the total number of energy levels in a finite sample of linear size $L$, 
 $|\psi_n ({\bf r})|^2$ is the probability density of the eigenstate
 at  the site ${\bf r}$ of the spin-$1/2$-impurity. 
 In a clean metal the Eigenstates are plane waves with
  uniform intensity $|\psi_n ({\bf r})|^2 = 1/L^d,$  
  The  disorder results in wavefunction amplitudes which vary with the  position ${\bf r}$. 
           In a weakly disordered metal  different 
           wave functions  
           are  correlated with each other in  
    a macroscopic energy interval of the order of  
     the elasic scattering rate $1/\tau$. Thus,  in a weakly disordered metal
     the Kondo  temperature  is found to have a  finite width in the thermodynamic limit \cite{jetpletters,micklitz05,prb2007}.
  
However, at the AMIT the wave functions are strongly inhomogenous and 
             multifractal \cite{multifractal} with power law 
             correlations in energy \cite{powerlaw,cuevas}.  
              Since the AMIT is a 2nd order quantum phase transition    
   the localization length  $\xi$ and the 
     correlation length  $\xi_c$  diverge at the AMIT. 
     Thus,  the electrons at the AMIT are in a critical state, 
  which is neither extended nor localized. In fact it is rather sparse,  see  
 Fig. \ref{mf},   where  the   intensity $|\psi_n({\bf r})|^2$  is plotted  for all sites of a 
 finite sample of  
 the 3D Atb model. 
At the AMIT the $q$-th  moment of eigenfunction intensities $|\psi_l({\bf r})|^2$ is found to 
scale with   linear size $L$ as
\begin{equation}
\label{single}
P_q = L^d \langle\, |\psi_l({\bf r})|^{2 q} \rangle \sim L^{-d_{q} (q-1)}.
\end{equation}
   Critical states are  characterized by 
    multifractal dimensions $d_{q} <d$  which change with   power $q$.
The  resulting distribution function of the intensity
is for the 3D AMIT very  close to  a log-normal distribution function \cite{Evers2008,rodriguez2011}
\begin{equation}
\label{eq:Pone}
P(| \psi_l({\bf r}) |^2) \sim \frac{1}{| \psi_l({\bf r}) |^2} 
L^{ - \frac{(\alpha_{\psi}-\alpha_0)^2}{2   \eta  } }.
\end{equation}
Here, $ \alpha_{\psi} = - \ln | \psi_l({\bf r}) |^2 / \ln L $, $  \eta  = 2 (\alpha_{0}-d)$ and 
$\alpha_0>d$.   The multifractal dimension $d_{q}$ is 
 related to $\alpha_{0}$ by $d_{q}= d - q (\alpha_{0}-d) $
  for  $q< q_c$.  
  At $q_{c}= \alpha_{0}/\eta$
  the function   $\tau_{q} = d_q (q-1)$ terminates at 
  $  \tau_{q_{c}}$ and does not change for  larger moments,  $q>q_{c}$ \cite{Evers2008}. 
Approaching  the AMIT,
         single particle wave functions 
           show multifractality when looking at  
           length scales  which are smaller than  $\xi$, $\xi_c$,
           respectively \cite{multifractal}. 
            Another consequence of multifractality is that 
             the intensities are power law correlated in energy \cite{powerlaw,cuevas,ioffe}, see Fig. \ref{mfc} and in space \cite{Evers2008}.
       The correlation function of 
intensities associated to two energy levels  is found to decay with a power law of their difference
$\omega_{nm} =E_n- E_m$ as
 \cite{cuevas,ioffe} 
\begin{eqnarray}
\label{cc}
  C(
\omega_{nm} = E_n-E_m) = L^d \int d^dr\, \langle |\psi_n({\bf r})|^2
 |\psi_m({\bf r})|^2 \rangle 
 \nonumber  \\
 = \left\{
 \begin{array}{ll}(\frac{E_c}{{\rm Max} (|\omega_{nm}|, \Delta) })^{\eta/d}, 
 & 0 < |\omega_{nm}| < E_c, \\ (E_c/|\omega_{nm}|)^{2}, &
|\omega_{nm}| > E_c,
\end{array}
\right.,
\end{eqnarray}
when $E_n \le E_M$ and $E_m \ge E_M$ or  the other way around, as plotted in   Fig. \ref{mfc}.    $\Delta = 1/(\rho L^d)$ is 
the average level spacing,  $\rho$  the average density of states. 
Here, we set one of the energies at the mobility edge $E_n=E_M$
 and the other at $E_m=E.$
   The correlation energy $E_c$ is  found to be  of the order
  of   elastic scattering rate $1/\tau$.
For
$|\omega_{nm}| < E_c$ correlations are enhanced compared to the
plane wave limit $C_{nm}=1$, see Fig. \ref{mfc}. 
Recently, it was  shown  in  numerical simulations \cite{harashima2014,amini,Carnio2019b} 
and 
 using the nonlinear sigma model  of 
  disordered interacting fermions \cite{burmistrov,burmistrov2} that  multifractality
persists   in strongly  interacting disordered fermion systems
 with {\it  long range  interactions}. 
 In cases when the interacting states cannot be written as Slater determinants of single particle states,  one can formulate the problem in terms of local density of states (LDOS).
 In Ref.    \cite{burmistrov2}  the LDOS   at the MIT was studied
 and confirmed to have a multifractal distribution.

             \begin{figure}[h]
\includegraphics[width=0.5\textwidth]{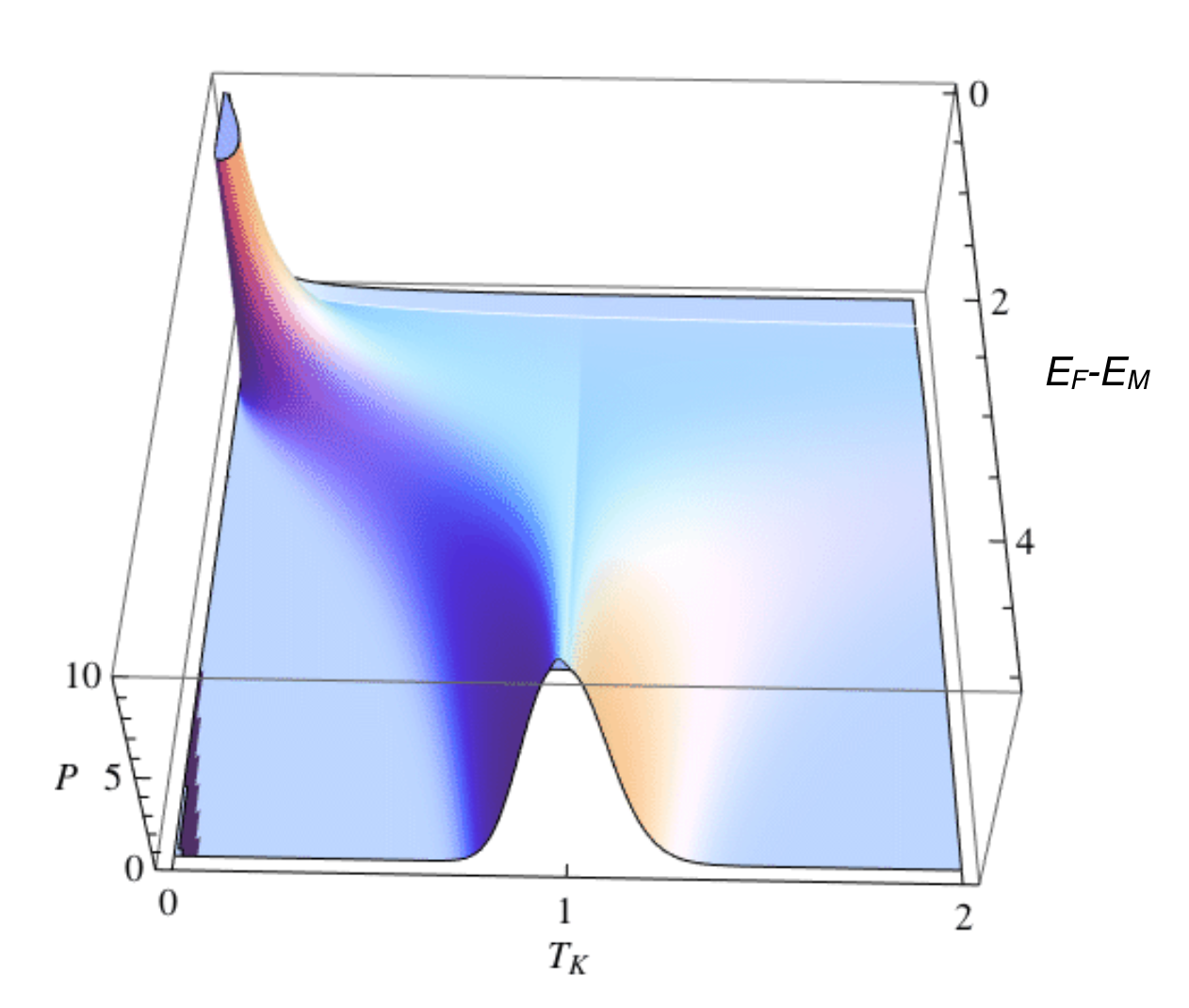}
\caption{ Distribution of Kondo temperatures $T_{K}$ in units of
  $T_{K}^{(0)}$ in the metallic phase, 
 acc. to  analytical theory Eq. (25) of Ref.  \cite{Kats}, 
 as function of  distance to mobility edge
  $E_{F}-E_{M }$ in units of $E_{c}$ for exchange coupling $J= D_0/5$, $D_0$ band width.
    Fig. taken  from 
 Ref.   \cite{Kats},  Copyright  2012, American Physical Society.}
\label{fig:ptkmetal}
\end{figure} 

    \begin{figure}
\begin{center}
\includegraphics[width=0.4\textwidth]{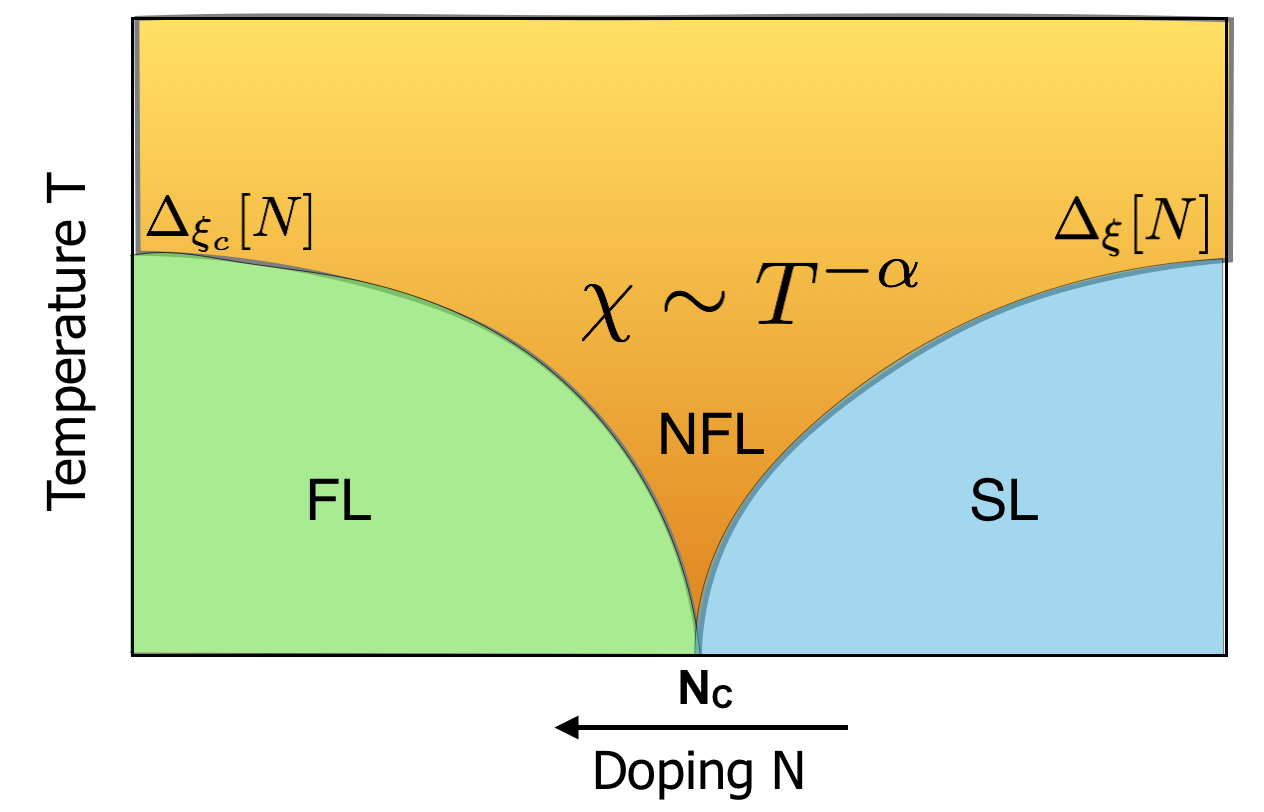}
\caption{Schematic phase diagram as function of doping concentration $N$ and temperature $T$. Anomalous magnetic susceptibility $\chi \sim T^{-\alpha}$ with universal power
  $\alpha = 1- (\alpha_0-3)/3$
at criticality $N_c$ and 
 for $T> \Delta_{\xi}$  in the insulator regime $N< N_c,$  and  for $T> \Delta_{\xi_c}$  in the metal  regime $N> N_c.$ 
}
\label{fig:fan}
\end{center}
\end{figure}

  Implementing these power law correlations of local wavefunction intinsities, 
  an  analytical result for 
              the distribution of $T_K$ at and in the vicinity of the AMIT 
              has been obtained in Ref.  \cite{Kats},     
            using that the  distribution  of wave function intensities 
         is close to  log-normal. 
                That distribution of $T_K$ is found to 
              widen as  
             the  system approaches the AMIT from the metallic side,
              evolving into a distribution with a power law divergent
              low $T_K-$ tail. That tail has been shown 
              in Ref.  \cite{Kats} to origin from 
               the opening of  local pseudo gaps, which quench the Kondo screening.
                The full distribution of $T_K,$
             Eq. (25) of Ref.  \cite{Kats}, is plotted
               in Fig. \ref{fig:ptkmetal} as function of the distance to  mobility gap $E_M$, approaching it 
               from the metallic side.
               At the mobility gap $E=E_M$ its tail is found to be given by 
               \begin{equation}
               P(0< T_K \ll T_K^0) \sim T_K^{-\alpha},
                \end{equation}
               with power $\alpha = 1- \eta/(2d),$
              depending only on the 
         multifractal correlation  exponent $\eta= 2(\alpha_0-d)$.     
          In the vicinity of the AMIT that universal power law tail remains     
           valid on the metallic side  $N> N_c$ for $T_K>\Delta_{\xi},$ 
           and 
            on the insulating side $N< N_c$ for $T_K>\Delta_{\xi},$   
                 where only at lower temperatures the Kondo screeing is fully quenched by local gaps 
                $ \Delta_{\xi}$ and a finite density of free magnetic moments remain. 
              This analytical result for the  distribution function of $T_K$  
              allows to calculate the  magnetic susceptibility
              noting that $ \chi (T) \sim n_{FM}(T)/T $ with $n_{FM} (T) = n_M \int_0^T d T_K  P( T_K)$,
              yielding \cite{Kats},
            \begin{equation}
            \chi (T) \sim  
   \left(\frac{T}{E_{c}} \right)^{-\alpha},
  \end{equation}
valid  at the MIT $N=N_c$, on the insulating side $N< N_c$ for $T>\Delta_{\xi}$ and  on 
the metallic side $N< N_c$ of the MIT for $T>\Delta_{\xi_c}$. This results in the fan like phase diagram, shown schematically in 
 Fig. \ref{fig:fan}. 
  The contribution of MMs to the specific heat  is obtained  
   from $C( T)  \sim T d n_{FM} (T)/dT$ as $C(T) \sim T^{\eta/(2d)}$. 
   In $d=3$ 
  we thus find that the  anomalous power of the magnetic susceptibility $\alpha$ has a universal value, independent of doping concentration $N$, as given by
   \begin{equation} \label{universal}
   \alpha = 1- (\alpha_0-3)/3,
   \end{equation}
    in the vicinity of the MIT. 
   The multifractal parameter  $\alpha_0$ is 
   without magnetic field known numerically   to be  $\alpha_0  = 4.048(4.045, 4.050) $ \cite{rodriguez2011}, 
    yielding  $ \alpha = .651(.652,.650)$.
    Thus, the critical  power law tail of the distirbution of the Kondo temperature 
    results in  an 
     anomalous power for the temperature dependence of the 
      magnetic susceptibility which is  in  good agreement with experiments in the  vicinity of the MIT, see the  inset (circles)  of 
   Fig. \ref{suscsip} of Ref.   \cite{Schlager1997} and results reported in Ref.     \cite{bhatt88}.  If this  interpretation is correct, it   would make these experiments 
    the first, albeit indirect, measurements of multifractality at the MIT in doped semiconductors. 
   
   A direct numerical calculation of the distribution of the Kondo temperature, 
   by application of  the
   Kernel polynomial method \cite{Weisse} to  Eq. (\ref{eq:FTK})  was reported  in  Ref. 
     \cite{Slevin19}.
    In the Atb model with $D=12t$   in $d=3$ at the AMIT  for $J = 4t$  
    a low $T_K$ divergence corresponding to 
  the anomalous power of the magnetic susceptibility of 
  $\alpha \approx   0.75$  was found  \cite{Slevin19}. This value is 
   very close to the one obtained by numerical exact diagonalisation 
    at the AMIT of the three dimensional Atb  in Ref.  \cite{cornaglia}.
    The remaining  discrepancy with the analytical value can be attributed to the two main 
     approximations employed in the analytical derivation: 1. parabolicity of the multifractal spectrum 
      $f(\alpha)$, and 2.  taking into account only pair correlations between electronic intensities at different energies. 
     The multifractality parameter 
     $\alpha_0$ was found numerically to change
      in   presence of MMs for the $d=3$ Atb model 
      to $ \alpha_0 =  4.53 \pm .07$\cite{Jung2016}. 
Inserting this value   into 
Eq. (\ref{universal}) yields for 
 the anomalous power of the magnetic  susceptibility 
$\alpha =.49 \pm .02,$
 improving the agreement with the measured values, given in the inset of Fig.  \ref{suscsip}, further.
      
   Going beyond the 1-loop equation  for the Kondo temperature,   Eq. (\ref{eq:FTK}), 
    it was found in Ref.  \cite{Zhuravlev2007} that  the distribution obtained by  a full Wilson renormalization group (RG) calculation   \cite{Wilson75,bethe}   for a disordered system yields a qualitatively similar distribution function. 
    Recently, it was pointed out that a more realistic  model of the Kondo impurity
    which takes into account anisotropies may 
    yield a modified distribution of  Kondo temperature   \cite{Debertolis2022}. It remains to be explored whether this will affect the low $T_K$ tail at the AMIT and thereby the anomalous  power
    of the magnetic susceptibility  $\alpha.$

  \section{  Coupling between Magnetic Moments, Spin Liquids}
   Even though there is  thus convincing evidence that  the  distribution of 
    Kondo temperature is dominating 
     anomalous magnetic properties close to the MIT in doped semiconductors, 
     for uncompensated Si:P  in 
     a temperature range of $10 {\rm mK } < T < {\rm 1 K}$ with an average Kondo temperature 
      of about $\langle T_K \rangle \approx 0.8 {\rm K}$  \cite{lakner}, 
       there  is also evidence that there remains a finite density of 
      free MMs on the insulating side of the transition in the low temperature limit,
    since the Kondo screening becomes quenched by  Anderson localisation, 
   where the renormalisation of the Kondo coupling becomes cutoff by the 
    local level spacing $\Delta_{\xi} = 1/(\rho \xi^3)$ \cite{meraikh}. 
     Therefore, at temperatures $T <  \Delta_{\xi} $, see Fig. \ref{fig:fan},
      a finite 
    density  of free MMs  remains
  which in the zero temperature limit is found to be  given by 
    $n_{\rm FM}(T=0K) = n_{M} \xi^{-\frac{ 1}{2   \eta } (d j)^2}$ \cite{Kats}.
   In a system of finite linear size $L$ it is  found that   
   at the AMIT the concentration remains finite, given by   $n_{\rm FM}(T=0K) = n_{M} L^{-\frac{ 1}{2   \eta } (d j)^2},$
    vanishing only in the thermodynamic limit. 
     The numerical finite size  calculations in Ref. 
     \cite{Slevin19} found even a  much larger concentration of free MMs at the  AMIT.   
    In Ref.   \cite{BhattFisher92} it was found  
     that even on the metallic side of the MIT there can remain free MMs due to 
     rare sites which due to the random placement  of dopant sites remain strongly isolated. 
     
    
    \subsection{Random Singlet State}  Since these MMs  are still 
      weakly coupled to the electron system, which mediate the exchange coupling, the MMs are  coupled with each other.
          In the dilute 
                 limit it is the  super-exchange  coupling
                  as  derived from the overlap of   hydrogen like impurity states between neighboured dopants, which
is known to be antiferromagnetic.
 The  randomly positioned MMs have therefore  been modeled 
by a Heisenberg spin model with random antiferromagnetic
   short range, exponentially decaying super-exchange   interaction 
   \cite{Andres1981,bhattlee81,bhattlee82,Paalanen1988,Sachdev1989}.
  In agreement with  experiments, numerical simulations and the implementation of  a
  cluster renromalisation group by Bhatt and Lee for  
  this model have  found no evidence of a  spin glass transition, 
  at which the magnetic susceptibility would peak and then decay to lower temperatures \cite{bhattlee81,bhattlee82}.
  Rather,  the 
  magnetic susceptibility of that model diverges at low temperature with a power law
  $\chi (T) \sim T^{-\alpha},$ with  $\alpha \le 1$ \cite{bhattlee81,bhattlee82}.
    In one dimension, the random antiferromagntic short range Heisenberg model  is known to 
     flow at low temperature to the infinite randomness fixed point, where 
    the ground state  is formed of  randomly placed  spin singlets, the so called {\it random singlet phase}.  This can be derived by  the strong disorder RG (SDRG) method. The temperature dependence
     of its magnetic susceptibility  is found to diverge almost Curie like \cite{bhattlee81,bhattlee82},
     and can be derived analytically in 1D,  yielding only  logarithmic corrections to the Curie law as 
      $\chi (T) \sim T^{-1}/\ln^2 (\Omega_0/T)$
      \cite{igloimonthus}.                         
                 
                  \subsection{Large Spin Fixed Point} 
   When the localisation length $\xi$ exceeds the Fermi  wave length $\lambda_F$, however,  
      the  interaction  between spins is longer ranged and oscillates in 
     sign with distance, similar to the  RKKY interaction
        in the metallic regime  \cite{rkky}, but decaying exponentially  beyond the 
        localisation length $\xi$.    
A short-range  Heisenberg model  
 with random sign coupling was studied 
using  a  modified version of the SDRG method in Refs.  \cite{sigrist,sigrist2}:  If the strongest bond is AF   it forms a singlet and adjacent spins  are coupled
             by  renormalized  interactions.
  If the strongest bond is FM  the Hamiltonian  is rather  projected onto 
    triplet states with reduced couplings.
           For  chains it was thereby shown that for any finite number of FM couplings 
       the chain is driven to a new fixed point with  clusters forming  large effective spins, contributiong  a  Curie law magnetic susceptibility \cite{sigrist2,Wan02}.        In higher dimensions it is known that even if the initial distribution is purely antiferromagnetic, ferromagnetic couplings can be generated upon renormalization \cite{bhattlee82,Motrunich00}. 
       
       Thus, the low temperature magnetic properties 
    of doped semiconductors is 
    expected to be dominated by  random singlets in the dilute limit, while at larger doping on the insulating side of the transition, clusters of larger effective spins may form. In both cases, only small corrections to the  
     Curie law are expected in the low temperature limit, where the MMs form a spin liquid,  and no spin glass formation is expected,  
     as indicated in the 
      phase diagram Fig. \ref{fig:fan}. 
          
            On the 
              metallic side of the MIT 
               the Kondo screening of the  MMs competes with  
              long range indirect  exchange  interactions.
              For a pure metal these are 
       known as 
               RKKY interactions \cite{rkky}. 
            Close to the AMIT their  amplitude is  however widely distributed 
              with a lognormal distribution \cite{lerner}, as can be understood as follows.   
   Writing 
 the indirect exchange
coupling between MMs 
 in terms of  local
 wave functions $ \psi_{n}({\bf r}_{i})$ one gets
\begin{eqnarray}
\label{eq:Jij}
 J_{{\rm RKKY}}({\bf r}_{kl})  = \frac{S(S+1) J_{ik} J_{jl} V_{0}^2}{4 \pi S^2} 
 \times \nonumber \\
\Im \int_{-\infty}^{E_{\rm F}} d \omega \sum_{n,l}
\frac{\psi^*_{n}({\bf r}_{i}) \psi_{n}({\bf r}_{j}) }{\omega - E_{n} +i \epsilon }
\frac{\psi_{l}({\bf r}_{i}) \psi^*_{l}({\bf r}_{j}) }{\omega - E_{l} +i \epsilon
},
\end{eqnarray}
where $V_{0}=L^3/N$.
 We see, that it depends not only on  intensities $ |\psi_{n}({\bf r}_{i})|^2,$ but also on the phase of the Eigenfunctions.
  This complicates its evaluation, especially at the MIT.  
In a  clean metal  the  RKKY-expression   is recovered when  inserting plane-wave
states
$\psi_{n}({\bf r}_{i}) \sim \exp (i {\bf  k r}_i)$  
 into Eq. (\ref{eq:Jij}). This  gives  for $k_{\rm F} r_{ij} \gg 1,$ $
 J^{0}_{{\rm RKKY}}({\bf r}_{kl})
\sim   J_{ik} J_{jl} \cos (2 k_{\rm F} r_{kl})  r_{kl}^{-3}$. 
Disorder  shifts the phases of  wave functions randomly.
This results in exponential 
suppression of  the couplings, giving
  $\langle J_{\rm RKKY} \rangle \sim \exp (-r_{kl}/l_{e})$
 when averaged over an ensemble of disordered
systems, where $l_e$ is the elastic mean free path  \cite{degennes}. 
       The  typical value  $ \sqrt{\langle
  (J_{\rm RKKY})^2 \rangle}$ is however found to remain  close to
$J_{\rm RKKY}^{0}$ for weak disorder.  
         At strong disorder 
         and at  the AMIT  the  electron intensities  $ |\psi_{n}({\bf r}_{i})|^2$
         have a lognormal distribution, as reviewed above. Thus, since  Eq. (\ref{eq:Jij}) is proportinal 
          to the intensities at two positions, one arrives at a 
          lognormal distribution of  $J_{{\rm RKKY}}({\bf r}_{kl})$ \cite{lerner}.
          This was  confirmed numerically
           for a 2D disordered system and for graphene  in Ref.  \cite{Lee2014}.  

        \subsection{\ Strong Disorder Fixed Point} 
        Random quantum spin systems  with long range  exchange interactions
         have recently been studied    by extending 
          the strong disorder RG method  to long range coupled transverse Ising chains
           \cite{Juhasz2014} and to 
           the random  long range coupled antiferromagnetic 
           XX-quantum spin chain  \cite{Moure2015,Moure2018,Mohdeb2020,Juhasz2020,Mohdeb2022}.
           Comparison with numerical exact diagonalization and tensor network extensions of the 
            Density Matrix RG method  \cite{Mohdeb2020,Mohdeb2022} confirmed   that 
            for power law interactions decaying with distance $R$  between the spins as
            $R^{-\beta}$  these models flow 
           to a  strong disorder fixed point, a random singlet state. The fixed point distribution of 
            the exchange couplings has a finite width ${\sl W} = 2 \beta$
           and the power law divergence of the  magnetic susceptibility is found to  depend
            on $\beta$ as
          $\chi(T) \sim T^{1/\beta-1}$ for $\beta >1$  \cite{Moure2018,Mohdeb2020}.
            At  $\beta =1$  there is a delocalization transition and $\chi(T)$ is expected to diverge logarithmically, only. 
            
         \subsection{SYK-model} 
              The random sign coupling
              was taken into account in an SU(M)-model with infinite range interaction whose 
              amplitude  is normally distributed 
            \cite{Bray80,SachdevYe1993},
           which is related to the SYK-model  \cite{Kitaev2015,Sachdev2015,Gu2020,SachdevRMP}.
            Due to its infinite range interaction, nonperturbative results can been obtained. Perfoming 
             the disorder average by means of the  replica  trick \cite{Bagrets2017} or the 
            supersymmety method  \cite{Tigran2020}  
             it can be mapped on a zero-dimensional  nonlinear sigma model.   
           For large $M>>1$ its ground state was shown to be  a spin liquid
           and the  magnetic susceptibility was found to have a 
            logarithmic divergency $\chi(T) \sim |\ln T|$ \cite{Bray80,SachdevYe1993},
           similar to what had been conjectured for the magnetic susceptibility
            in marginal Fermi liquids  \cite{Varma1989}.
           For spin $S=1/2$, corresponding to $M=2$, the ground state was recently found 
           to show spin glass type correlations
             \cite{Christos2022}, similar to the corresponding random classical Ising model 
            with infinite range interaction \cite{Sherrington1975,fisch1980}, 
            but excited states were found in Ref.   \cite{Christos2022} to show spin liquid behavior. 
         
     \subsection{More Realistic Models of Randomly Coupled Quantum Spins}    To get a better understanding of the magnetic properties of doped semiconductors at their MIT, 
             power law long range coupled  Heisenberg spin $S=1/2$-models with random sign and
             wide, lognormal distribution of their exchange couplings still need to  be studied in 
             $d=3$.       
            Such studies have been  mentioned in Ref.  \cite{bhattlee82} where no spin glass ordering has been found for power law couplings with power $\beta \ge 3$. 
            
            Recently,  it became possible to explore   disordered spin ensembles at a diamond surface   by probing 
          it with  single nitrogen-vacancy (NV) centers in diamond
               \cite{Lukin2022},   
   \cite{Davis2022}.
   Recent advances in experimental 
setups in cold atom systems 
allow the  study  of long range coupled spin systems  with interactions that fall-off as $1/r^3,$
which has been demonstrated by coupling  Rydberg states with opposite parity \cite{Signoles2021,Brow2020,Franz2022}. Trapped ions with power-law interactions, decaying as $1/r^\alpha$, with tunable $0<\alpha<1.5$  have also  been realized\cite{ex6,ex7,ex8}, recently.
    As these setups allow the  controlled experimental study of
   real  systems of randomly coupled quantum spins, they 
   may allow analogue simulations of   models and 
  promise to contribute to a better understanding of 
   the magnetic properties of doped semiconductors.

   \section{Competition between Kondo Effect, Indirect Exchange Interaction and Disorder}
  
  \subsection{Doniach Diagram} 
   The indirect exchange interaction of
     MMs competes with   the Kondo effect as 
   the local exchange coupling $J$ and the concentration of MMs $n_{M}=  R_{}^{-d}
$  is varied, where    $R$ is
the average distance between neighboured MMs, 
 This  gives rise to a rich quantum phase diagram, the Doniach diagram
   \cite{Doniach,Kroha17}, where 
a Kondo screened phase is separated from a  phase which is dominated by the 
   indirect exchange interactions.
    There, 
below a   critical density $n_{c},$  the Kondo effect is
prevailing the RKKY interaction.
For  an  electron  system without  nonmagnetic disorder that critical density  is found  from the condition 
$|J^{0}_{\textrm{RKKY}} (R_c) | = T_K$.
For example, in a 3D sample with
$|J^{0}_{\textrm{RKKY}}|_{k_F R \gg 1} = J^2 \rho(\vep_F) \frac{\cos(2 k_F R)}{32 k_F^3 R^3}$ and
$T_K = c \vep_F \exp (- 1/(\rho J))$, where $k_F$ is the Fermi momentum and $c \approx 1.14$, 
one finds the  critical electron density 
$  n_c = 32 c \frac{\vep_F}{\rho J^2}  \exp (-\frac{1}{\rho J})$, increasing with $J$ 
in the relevant coupling range $J \rho <1$.
%
 In a disordered system  
      the  Kondo temperature $T_{K i}$
     is different at each  site ${\bf r}_i$ and competes with  the RKKY coupling
      $J_{\textrm{RKKY}}({\bf R}_{i j})$
      with other MMs  which are located at  sites
      ${\bf r}_j$  at  distance ${\bf R}_{i j}$.
    Their ratio \,$J_{\textrm{RKKY}}({\bf R}_{i j}) / T_{K i}$,
varies  across the  disordered
sample.  This problem has been studied with a disordered  Kondo lattice model 
   \cite{Coqblin,Burdin2007,Burdin2021}
and  with the Anderson-Hubbard model with  numerical methods, 
 including dynamical mean field theory based approaches  \cite{Byczuk2009,Ulmke1995,Aguiar2006,Aguiar2009,Kotliar2003,Aguiar2013,Byczuk2005,Weh2021}, and Hatree-Fock based approaches  \cite{Milovanovic1989,Sachdev1998,Tusch1993}, quantum Monte Carlo  \cite{Byczuk2011,Ulmke1997,Pezzoli2010}, 
 and  the typical medium dynamics cluster approximation  \cite{Jarrell2017,Jarrell2014}.
              In Refs.  \cite{Jarrell2017,Jarrell2014}
 the quasiparticle self energy  of the Anderson-Hubbard model  was
  derived as function of excitation energy $ \omega,$
     $Im \Sigma ( \omega) \sim \omega^{\alpha_{\Sigma}}.$
 It was 
   found to  behave  as a non-Fermi liquid   with power $\alpha_{\Sigma} (W) < 2,$
 which was found to  become smaller with
      stronger   disorder amplitude $W$. 
     The non-FL is found to extend down to the lowest energy
      at the MIT,   while there
     is a crossover away from the MIT, 
    as limited by  a cutoff  which coincides with  the local level spacing in the insulator phase,
         $\Delta_{\xi} =1/(\rho \xi^d)$, and  $\Delta_{\xi_c} =1/(\rho \xi_c^d)$, in the metal phase, respecively. 
         This is  in agreement with the phase diagram derived from the magnetic properties  and reviewed above, as shown in Fig. \ref{fig:fan}. 
        Ref.    \cite{Jarrell2017} also reported 
 the distribution of Kondo temperatures, as defined by 
     the width of the spectral function peak, and found 
      that it widens at the MIT into a power law tail, qualitatively very similar
       to our analytical result, shown in  Fig. \ref{fig:ptkmetal}. 
      Since the  typical medium dynamics cluster approximation does not  model  indirect exchange interactions,  the Doniach diagram was not studied in Ref.    \cite{Jarrell2017}. 

\begin{figure}[h]
\begin{center}
\includegraphics[width=0.4\textwidth]{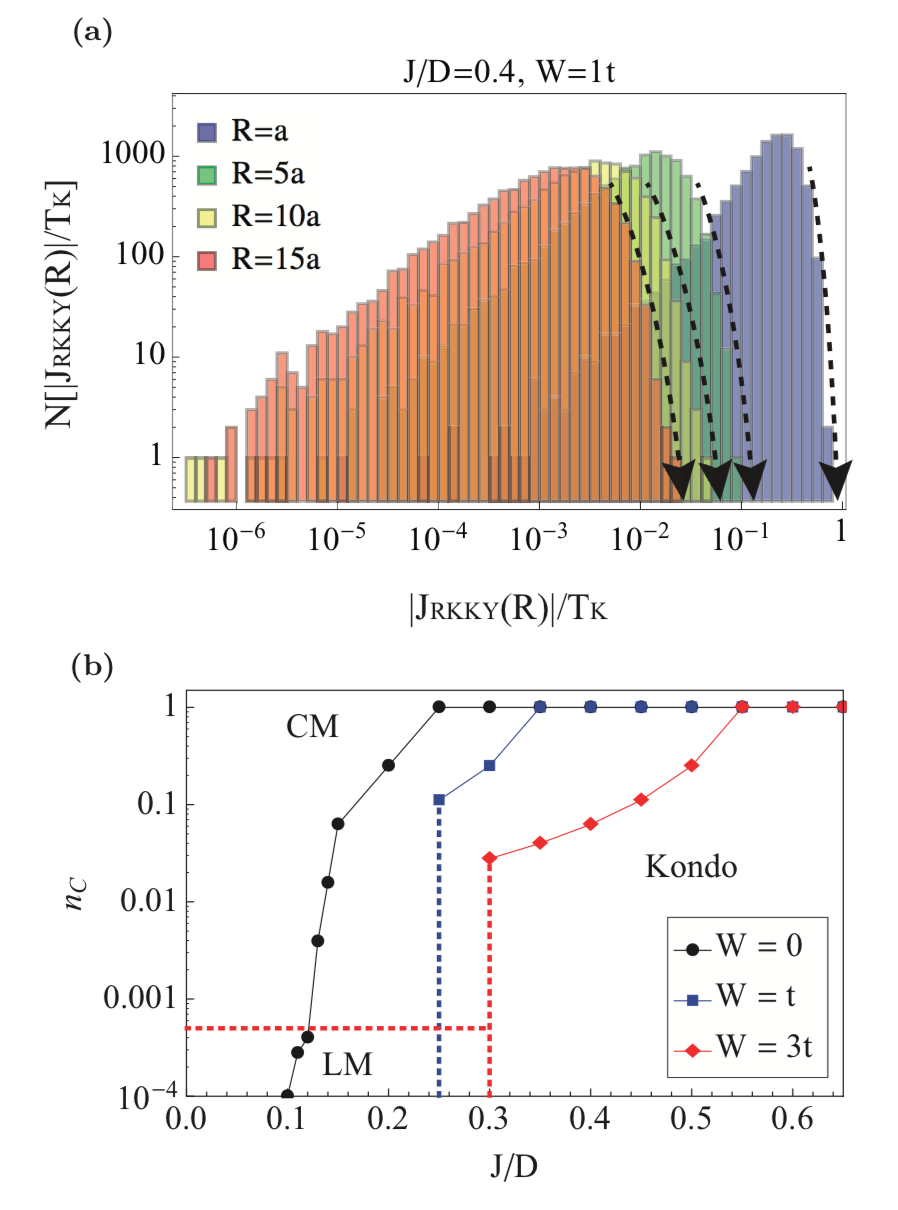}
\caption{ (a) Distribution of the ratio between $|J_{RKKY}(R)|$ and $T_K$
 for various distances $R$ between two arbitrary MMs for a 2D Tbm. Black dashed arrows highlight the sharp cutoff for  each distribution. (b) Magnetic quantum phase diagram: critical MM density $n_c$ (as determined by  the distance $R_c$ such   $|J_{RKKY}(R)/T_K|<1$   
 for all sites when $n_M = 1/R^d < n_c = 1/R_c^d$)  as function of $J /D_0$ ($D_0$  is the band width) for various disorder strengths $W.$  Horizontal dashed line separates the   local moment phase (LM). Linear system size is  $L = 100a$, the number of disorder configurations  is $30 000.$.
  Fig. taken  from 
 Ref.    \cite{Lee2014},  Copyright  2014, American Physical Society.
}
\label{fig:ratio}
\end{center}
\end{figure}

            In order 
 to get a better understanding of the Doniach diagram in disordered systems
 the  distribution of the ratio  between  $J_{\textrm{RKKY}}({\bf r}_{i j})$
and $T_{K i}$ has been calculated in Ref.   \cite{Lee2014}, as shwon  in  Fig. \ref{fig:ratio} a) for  a 2D Atb model
 for four distances $R$ between two arbitrary MMs. The  black 
 dashed arrows highlight a sharp cutoff  found for these  distributions.  In  Fig. \ref{fig:ratio} b) we show the  critical MM density $n_c(J)$ 
  (as determined by  the  distance $R_c$, which is defined by the condition    $|J_{RKKY}(R)/T_K|<1$  for all
  $n_M = 1/R^d < n_c= 1/R_c^d$)
   as function of exchange coupling  $J$
 in units of    band width $D_0$,  for three disorder strengths $W$.
 At strong disorder, the RKKY coupling is exponentially cutoff by the localisation length, yielding 
 at small  densities a   local moment phase (LM). 
  While such a study remains to be done in 3D disordered systems at the AMIT, 
   we can deduce from the study in 2D, Fig. \ref{fig:ratio}, that the 
    coupling between MMs becomes more likely to dominate the Kondo screening as the MIT is approached, since 
     the density of  MMs and the disorder strength increases as the doping density is reduced towards the MIT. 

  \subsection{Selfconsistent Renormalisation Group  Theory of Disordered Kondo Lattices} 
Including  the RKKY coupling between MMs selfconsistently in the 
  renormalization group equations for  a  Kondo  lattice   \cite{Nejati2017}, 
   it was recently shown that the Kondo temperature  decreases
   when one increases  the RKKY coupling. This happens   before   the Kondo screening is  quenched  
   completely at couplings above  a critical RKKY coupling. Thus,   when considering a disordered system, 
  it is actually not sufficient to calculate $J_{\textrm{RKKY}}({\bf r}_{i j})$
and $T_{K i}$  independently, but the effect of $J_{\textrm{RKKY}}({\bf r}_{i j})$
on  $T_{K i}$ needs to be taken into account. 
 Let us therefore review this theory briefly. 
The effective Kondo coupling $g_i  = \rho_i(\mu) J_i$  of the Kondo impurity at site ${\bf r}_i$ 
with the density of states at the chemical potential per spin $ \rho(\mu)$
  is governed by  RG-equations as  modified  by  the presence of RKKY couplings  \cite{Nejati2017}
\bqa && \frac{d g_i}{d \ln D} = - 2 g_i^{2} \Big( 1 - y_i g_{0}^{2} \frac{D_{0}}{T_{K}} \frac{1}{\sqrt{1 + (D/T_{K})^{2}}} \Big) . \label{rgkj} \eqa
$D$ is the energy scale which  cuts off the RG-flow. While the first term  on  the right  side
 is known as    the 1-loop $\beta-$function of the Kondo effect, the second term  
 takes account of the RKKY
  coupling. Here, $g_{0} =\rho(\mu) J_0$
with  bare Kondo interaction $J_0$ and  bare bandwidth $D_{0}$.  The effective dimensionless RKKY coupling strength at site ${\bf r}_i$ is  defined by \cite{Nejati2017}
\bqa && y_i = - \frac{8  \mathcal{W}}{\pi^2 \rho(\mu)^2} {\rm Im} \sum_{j \neq i} e^{i {\bf k}_F {\bf r}_{i j}} G^R_c ({\bf r}_{i j}, \mu) \Pi ({\bf r}_{i j}, \mu).
\label{y}
\eqa
%
%
Here, $\mathcal{W}$ is the Wilson ratio, which is known from  
the Bethe Ansatz solution of the Kondo problem  \cite{bethe}. $G^R_c ({\bf r}_{i j})$ is
defined to be 
 the   conduction band  single particle  propagator from site ${\bf r}_{i}$ to ${\bf r}_{j}$.
 The summation runs over all other MMs at positions ${\bf r}_{j}$.
The RKKY  correlation function of  conduction electrons between sites  ${\bf r}_{i}$ and ${\bf r}_{j}$
is denoted here as  $\Pi ({\bf r}_{i j}, \mu)$. While the RKKY correlation function 
oscillates between  positive and negative values, 
  $y_i$ is  positive  \cite{Nejati2017}.
We note that the effective Kondo interaction $g_i$  is still  a function of $D/T_{K}$, 
even though the RG-equation includes the RKKY-coupling, when 
 $T_{K}$ is  understood to be the renormalized Kondo temperature. 
  $T_K$ has  to be found  self-consistently from that RG-equation (\ref{rgkj}).  

Considering first  two MMs in a clean system, with 
bare couplings $g_0$ and $y_i=y$,
the solution of the RG-equations  gives \cite{Nejati2017}
\bqa && \frac{1}{g} - \frac{1}{g_{0}} = 2 \ln \Big( \frac{D}{D_{0}} \Big) \nn \\ &&- y g_{0}^{2} \frac{D_{0}}{T_{K}} \ln \Big( \frac{\sqrt{1 + (D/T_{K})^{2}} - 1}{\sqrt{1 + (D/T_{K})^{2}} + 1} \Big) .  \eqa
When $D \rightarrow T_{K}$, one findst that the effective Kondo interaction $g$ diverges, $g \rightarrow \infty$. This condition gives a self-consistent equation for the effective Kondo temperature as a function of the RKKY-coupling,
\bqa && \frac{T_{K}(y)}{T_{K}(0)} = \exp\Big(- y k g_{0}^{2} \frac{D_{0}}{T_{K}(y)} \Big) . \eqa
Here, $T_{K}(0) = D_0 \exp (-1/(2g_0))$ is the bare Kondo temperature  without RKKY-coupling
and $k = \ln (\sqrt{2} + 1)$. 
The critical
 coupling at which the RKKY interaction  prevents  the Kondo screening completely, is given by   \cite{Nejati2017}
\begin{equation} \label{yc}
y_c= T^0_{K}/(k~ e g_0^2 D_0 ).
\end{equation}

Recently, this theoretical framework was  applied to two MMs with  
different  local density of states
    at different sites, giving 
  different bare Kondo temperatures,  $T^0_{K i} = D_0 \exp (-1/(2g^0_i))$ \cite{Park2021}.
  The resulting  coupled RG-equations were then solved  numerically.
 Thereby, it was found  that  both Kondo temperatures are reduced in the presence of 
 the RKKY-coupling. However, 
  the initially smaller Kondo temperature was found to be   suppressed more 
   than the initially larger one. Thereby,  their ratio $x = T_{K_{2}}/T_{K_{1}}$  decreases
  due to the RKKY-coupling.
  The smaller
  the ratio $x_0$ is, initially, the stronger the inequality $x$ becomes. 
  Thus,  inhomogeneity is  found to be a relevant perturbation. 
   Any inequality betwen  Kondo temperatures becomes enhanced   by the RKKY coupling.
Moreover,  the Kondo screening is destroyed already by smaller  RKKY coupling, 
  the stronger
  the inhomogeneity and the smaller the  initial ratio of bare Kondo temperatures  $x_0$ is. 
   Thus, one can conclude  that an 
  inhomogeneity makes 
   the Kondo screening of the magnetic impurities   more easily quenchable by RKKY coupling.
  
For a finite density of  randomly distributed MMs,  $n_M,$ which are coupled by
random
local exchange couplings $J^0_i$ to the conduction electrons with   local density of states $\rho(E, {\bf r}_i )$ one can extend this approach.
Every MM 
  has then, in general,  different Kondo temperatures, as they are placed at different positions, 
 yielding  a  distribution of Kondo temperatures  \cite{mott,BhattFisher92,Lakner1994,cornaglia,Kats,Slevin19}.
 As  reviewed  above, the RKKY coupling can  also have a wide distribution
  \cite{BhattFisher92,lerner,Lee2014}. 
The derivation of  the quantum phase diagram of
 a disordered electron system with  finite density of MMs $n_M$ remains therefore an open problem.
 With  the  indirect exchange  couplings Eq. (\ref{y})
we can formulate for 
the coupled, randomly distributed MMs 
the set of  self-consistent RG-equations.
Since  the   local density of states $\rho( {\bf r}_i,E )$ does depend on energy $E$
 at each  RG scale $D$
the  renormalisation of  local  exchange couplings  $ J(\bm{r})$
 depends on the local density of states $\rho( {\bf r}_i ,\mu \pm D)$.
 Since that 
  can   differ strongly  from the density of state at  chemical potential $\mu$, $\rho_{0 i} = \rho({\bf r}_i ,\mu)$, it is important to keep this energy dependence. 
We get for   $g_i = J(\bm{r}_i) \rho_{0 i}$  the RG-equations as
\begin{widetext}
\bqa \label{rgnm1} && \frac{d g_i}{d \ln D} = -  g_i^{2}
\sum_{\alpha=\pm} \left(
\frac{\rho(\mu+ \alpha D,\bm{r}_i)}{\rho_{0 i}}
- \frac{4 J_i^0}{\pi \rho_{0i}}
  \sum_{j \neq i} J_j^0 {\rm Im} [e^{i {\bf k}_F {\bf r}_{ij} } \chi_c ({\bf r}_{ij},\mu+ \alpha D)  G^R_c ({\bf r}_{ij},\mu+ \alpha D) \chi_f ( {\bf r}_{j},\mu+ \alpha D ) ] \right). \eqa
\end{widetext}
 While the first term on the right hand side is the well known  
  1-loop RG for the Kondo problem with energy  dependent  density of states,
 \cite{Suhl65,Zarand96,gapless},  the second term describes the inhomogenous coupling. 
Here, $\chi_f ( {\bf r}_{j},E) $ is
the  f-spin susceptibility of the MM
 at  site ${\bf r}_{j}$.
$ G^R_c ({\bf r}_{ij},E)$ is the retarded   conduction electron propagator at site 
 ${\bf r}_{i}$ to ${\bf r}_{j}.$ Here, we defined
${\bf r}_{ij} = {\bf r}_{i} - {\bf r}_{j}$.
$\chi_c ({\bf r}_{ij},E) $ 
denotes  the conduction electron correlation function between sites ${\bf r}_{i}$ and ${\bf r}_{j}$.
Solving Eq. (\ref{rgnm1})  we can thus derive   the position dependent Kondo temperatures  for  a given configuration of indirect exchange  interactions.

When the densities  of MMs $n_M$ is not too large,
 $\chi_f ( {\bf r}_{j},E) $  can be approximated  by the
 Bethe-Ansatz solution   for a single Kondo impurity \cite{bethe}. In
   Ref.  \cite{Nejati2017} this approximation has been used.  Then, 
only   its  real part contributes, as given by
  ${\rm Re} \chi_f ( {\bf r}_{j},\mu+D)  =  \mathcal{W}/(\pi T_{Kj} \sqrt{1+D^2/T_{Kj}^2}).$ Here, $\mathcal{W}$ is the Wilson ratio. $T_{Kj}$ is the Kondo temperature of the MM at site ${\bf r}_{j}$.
Since it is well known that
       the energy dependence of the density of states
        changes the Kondo renormalisation  \cite{gapless}
       and thereby  yields a different 
        Kondo temperature function  in disordered systems \cite{prb2007,Zhuravlev2007,Kats},
         it is important to keep the energy dependence of all functions and not to replace it with  their value at the chemical potential.

When knowing  the distribution of the
local couplings  $g_{0} (\bm{r})$ which originates from the random positions of  the MMs
 with
 random  distribution of
 the local density of states,
and  the long range
function $y(\bm{r}-\bm{r}')$,   we can thus
 derive  the  distribution function of Kondo temperatures $T_{K}$ from  Eq. (\ref{rgnm1}).
 We note  that the random distribution of RKKY-couplings is mainly
 due to the distribution of local couplings $g_{0} (\bm{r})$  \cite{Lee2014}.
 Therefore  the distribution of $T_K$  comes mainly from 
 local couplings  $g_{0} (\bm{r})$, while
 the function $y(\bm{r}-\bm{r}')$
  is only weakly affected  by the disorder.

As reviewed above, without the 
  RKKY-coupling  $T_K$ has close to the AMIT
 a bimodal distribution with  one peak close to the
   Kondo temperature of the clean system and the other peak at   low $T_K$   \cite{cornaglia,prb2007,Kats,Slevin19}.
For stronger disorder more weight is shifted to the   low $T_K$  peak.  
 At the Anderson MIT it converges to a universal power law tail, 
 as reveiwed above,  where the power exponent $\alpha$ has a universal value, which depends only on 
 the multifractality parameter $\alpha_0$, see Eq. (\ref{universal})  \cite{Kats,Slevin19}.
  Since the RKKY-interaction is found to 
   enhance inequalities between Kondo temperatures, it  is expected that these interactions 
   shift more weight to  the low Kondo temperature peak.
    This can be checked quantitatively  by the solution of    Eq. (\ref{rgnm1}).

  The interplay of the Kondo effect, indirect exchange interaction 
   and Anderson localization has recently been 
    studied in a 2D experimental setup in a controlled way \cite{Zhang2022}, which may open new
   research   directions towards a better  understanding of  this problem.

 \section{Effect of Magnetic Moments on the MIT}
   
    Having  established the presence of MMs in   doped semiconductors 
  across the MIT and   deep into the metallic regime, 
   the question arises, 
   how  MMs affect the MIT itself.
It is well known that Anderson localization is weakened by
MMs since their  coupling to  the spins of
 itinerant electrons breaks their time-reversal symmetry (TRS) and spin reversal symmetry (SRS) \cite{Hikami1980,Efetov1983,Efetov1997},
changing the universality class of the AMIT  
  from   orthogonal to   unitary 
 \cite{Khmelnitskii1981,Wegner1986}.
The AMIT in 3D disordered systems 
could  accordingly  be 
 affected by MMs,
 decreasing   the critical electron
density $n_{\rm c},$ increasing the critical disorder $W_c$, and changing the symmetry class of the transition
 \cite{Khmelnitskii1981,Wegner1986}. 
 
 \subsection{Scaling Theory of Anderson Localization in  an Orbital Magnetic Field}
   In  an external magnetic
field in 3D disordered systems this  crossover is known to  be governed by the parameter
 $X_B = \xi^2 / l_B^2$, with $l_B$ the magnetic length  \cite{Hikami1980,Khmelnitskii1981},yielding the   scaling Ansatz for
the zero-temperature conductivity in a magnetic field, 
$\sigma(B) = e^2 f(X_B)/(h \xi),$ which thereby   becomes  
a  function of the
disorder amplitude $W<W_c$  \cite{Wegner1986,Khmelnitskii1981}
\begin{equation}\label{swb}
    \sigma(W,B) =
       (W_c-W)^{\nu} \tilde{f}(l_B^{-2}  (W_c-W)^{-\varphi}) 
        \end{equation}
 with  the scaling function $\tilde{f}$. 
  This scaling Ansatz implies  that  the  critical disorder  $W_c(B)$ 
  (and respectively  the critical 
  density $n_c(B)$)
  depend on magnetic field $B$   \cite{Khmelnitskii1981}.   
 On the other hand, when coming from the metallic side of the transition $W < W_c(B),$
 Wegner scaling implies that 
   the  zero-temperature conductivity   scales  in $d=3$ as  
   $\sigma \sim (W_c(B) -W)^{\nu_B}$, where $\nu_B$ is the critical exponent in a magnetic field.
   Comparison with Eq. (\ref{swb}) thus gives 
   $W_c(B) = W_c + A l_B^{2/\varphi}$, $2/\varphi = 1/\nu$, when $\nu_B \approx \nu$. 
   This scaling of $W_c(B)$ with $\varphi = 2 \nu$ has been confirmed with 
    a numerical transfer matrix method in 
Ref.  \cite{Droese1998}. 

\subsection{ Two-scale Localization}
 We note that this  scaling with  magnetic field 
$B$, outlined above, is not the only possibility.
 A.V. Kolesnikov and K. B. Efetov  found  in Refs.  \cite{Efetov1999,Kolesnikov2000}
  evidence for 2-scale localization, 
 where they calculated with the supersymmetry method the impurity averaged  spatial density-density correlation function  in a disordered wire in a magnetic field, and found two terms decaying   exponentially  with  two different  localization lengths,  the orthogonal $\xi_O$ (as obtained in disordered wires with TRS) and  the unitary 
 one $\xi_U$ (as obtained when the TRS is broken). Thus, they concluded that it is not the  localisation length which crosses over with the magnetic field between the orthogonal and unitary limit, but rather the weight of their contribution changes with the magnetic field. Whether this  effect is due to  two-scale  properties of individual wave functions, or rather due to a bimodal distribution function of the localisation length, remains unclear. Numerical calculations which addressed the same problem, albeit considering different properties, did not find indications of 2-scale localization in disordered quantum wires   \cite{Pichard1990,Schomerus2000,Weiss2001}. Other analytical approaches to this problem were 
 either heuristic   \cite{Bouchaud1991},  made assumptions about the scaling with magnetic field
 \cite{Lerner1995} or  extracted the magnetic field dependence of the 
localisation length from a  different correlation function, the autocorrelation function of spectral determinants    \cite{Kettemann2000}, which finds a  smooth magnetic field dependence  of the localisation length in  good agreement 
 with experiments \cite{Gershenson}. Since the supersymmetric calculations given in Refs.  \cite{Efetov1999,Kolesnikov2000} for the density correlation functions remain undisputed, and 
  the physical argument in favor of 2-scale localization is plausible, this problem  is even more relevant in higher dimensions:  In  2D the difference between orthogondal and unitary localisation length is exponentially large \cite{Abrahams79,Wegner79,Efetov1980,Hikami1980-2,Wegner1989,Kettemann2004,meraikh} and in 3D the critical disorder differs  between the orthogonal and unitary limit. Thus, it  is certainly a relevant question to ask, whether the obove scaling assumption remains valid in the crossover regime, or whether  there is a coexistence of 
   orthogonal and unitary physics. Then,    strongly localised states, which do not see sufficient magnetic field  and are thus governed by the orthogonal class, could coexist 
   with more extended states,
   which  see sufficient magnetic field to be in the unitary class,  resulting in  further increase of their  localisation length or even their delocalisation. While 
   this review reports  in the following results which are based on the  scaling of the conductivity with magnetic field, assuming that the localisation length is self averaging with a normal distribution,  the question  of two scale localisation raised by K. B. Efetov in quasi-1D-wires \cite{Efetov1999,Kolesnikov2000} still needs to be addressed in higher dimensions, and it is
   in particular  a worthwhile problem to study its consequences for the metal-insulator transition
   in 3D.

 \subsection{Scaling Theory of Anderson Localization with Magnetic Moments}
Noting that magnetic scattering  breaks both TRS and spin symmetry, 
changing thereby the  symmetry class of the transition
 \cite{Khmelnitskii1981,Wegner1986}, 
that crossover is governed by  the spin scattering rate from MMs
$\tau_s^{-1}$  through the  parameter $X_s = \xi^2 /
L_s^2$,  where $L_s = \sqrt{D_e \tau_s}$ is the spin relaxation length \cite{Hikami1980},
 $D_e =  v_F^2\, \tau / 3$ the electron  diffusion coefficient.  When $X_s \ge
1$ the electron spin relaxes before it covers the area limited by correlation length $\xi.$ 
Thus, the magnetic length  $l_B$  becomes
 in the presence of 
 MMs replaced by  the spin relaxation length 
 $L_s,$ yielding for   $W < W_c$ the zero-temperature  scaling function of the conductivity, 
      \begin{equation} \label{sws}
   \sigma(W,L_s) =
       (W_c-W)^{\nu} \tilde{f}(L_s^{-2}  (W_c-W)^{-\varphi_s}).
        \end{equation}
   However, it was found that  $\varphi_s$ is
 an  anomalous scaling exponent  with 
$
    \varphi_s \approx  2\nu + 3,  
$ as derived 
in second-order $d = 2 + \varepsilon$-expansion
within the N-orbital  nonlinear sigma model 
with  spin scattering  \cite{Wegner1986}.  
 Noting that Wegner scaling of 
   the  zero-temperature conductivity    in $d=3$  gives for $W< W_c(L_s)$, 
   $\sigma \sim (W_c(L_s) -W)^{\nu_s}$, where $\nu_s$ is the critical exponent with 
   magnetic scattering, we obtain by comparison with  Eq. (\ref{sws})
that the critical disorder is shifted to 
$
    W_c(L_s) = W_c(0) + W_c(0) A
  L_s^{-2/\varphi_s} 
$ when $\nu_s \approx \nu$ and  with  $  \varphi_s \approx  2\nu + 3$.  $A$ is a positive constant and $W_c(0)$ is the critical disorder strength without MMs. 
This  relation was confirmed with 
 a numerical analysis in Ref.  \cite{Jung2016},
  where  the  coupling of 
conduction electron spins $\vec{\sigma}_i$ to 
classical  spin vectors $\vec{S}_i$ with random orientation was modeled.
There, the  critical exponent was found to be reduced to  $\nu_s = 1.34 \pm 0.01$ for $J/t= 0.45$.  
   Also, the multifractality parameter 
     $\alpha_0$ was found to change
      in   presence of MMs in $d=3$ to $ \alpha_0 =  4.53 \pm .07$. 
Inserting this value   into 
Eq. (\ref{universal}) yields for 
 the anomalous power of the magnetic  susceptibility the value
$\alpha =.49 \pm .02,$
 in better agreement with the measured values, given in the inset of Fig.  \ref{suscsip}.
 
 However, the
   spin scattering rate
   $\tau_s^{-1}(T)$ 
 from Kondo impurities depends itself  on temperature $T$.
At high temperature 
 the MMs  with   spin $S$ 
 have   the magnetic relaxation rate   
 given by $1/\tau_{s} (T\gg T_K) = 2\pi n_{M} S (S+1)
j^2 \rho (\epsilon_F)$, where $j= J/D$. 
When magnetic impurities are dilute, the Kondo effect  screens the
impurity spin  for temperatures $T < T_K.$ This results in  a vanishing spin relaxation rate at zero
temperature. At finite
temperature the Kondo correlations enhance the spin
relaxation rate, which becomes maximal at $T_K$. In weak-localization experiments 
 a plateau in the   temperature dependence of the 
dephasing
time has been explained in terms of this peaked  temperature dependent spin scattering from Kondo impurities \cite{bergmann,mw00}, as 
numerically studied in  Refs.  \cite{zarand,micklitz05,micklitz07}.
The  following approximate expression is in good agreement with these results in all temperature regimes
 \cite{jetpletters}
\begin{eqnarray}
\label{eq:tauapprox}
\frac{1}{\tau_s^{(0)}} (T) & = & \frac{\pi\, n_m\, S(S+1)}{\rho} \left\{
\ln^2 \left( \frac{T}{T_K} \right) \right. \nonumber \\ & & \left. +
\pi^2 S (S+1) \left[ \left( \frac{T_K}{T} \right)^2 + \frac{1}{\beta}
  - 1 \right] \right\}^{-1}.
\end{eqnarray}
Here,  $\beta = 0.2$ \cite{zarand}. Thus, the 
temperature dependence of $1/\tau_s(T)$ scales in the dilute limit with 
the Kondo temperature $T_K$. Note that,
according to Eq. (\ref{eq:tauapprox}),
 the spin relaxation rate vanishes as
$T^2/T_{K}^2$ in the
low-temperature limit. Since this  has the same temperature dependence as
 the  inelastic scattering rate in a
Fermi liquid  this confirms   the  renormalized Fermi
liquid theory of dilute Kondo systems, as formulated by Nozi\` eres  \cite{nozieresfl}.

\begin{figure}
\begin{center}
\includegraphics[width=0.45\textwidth]{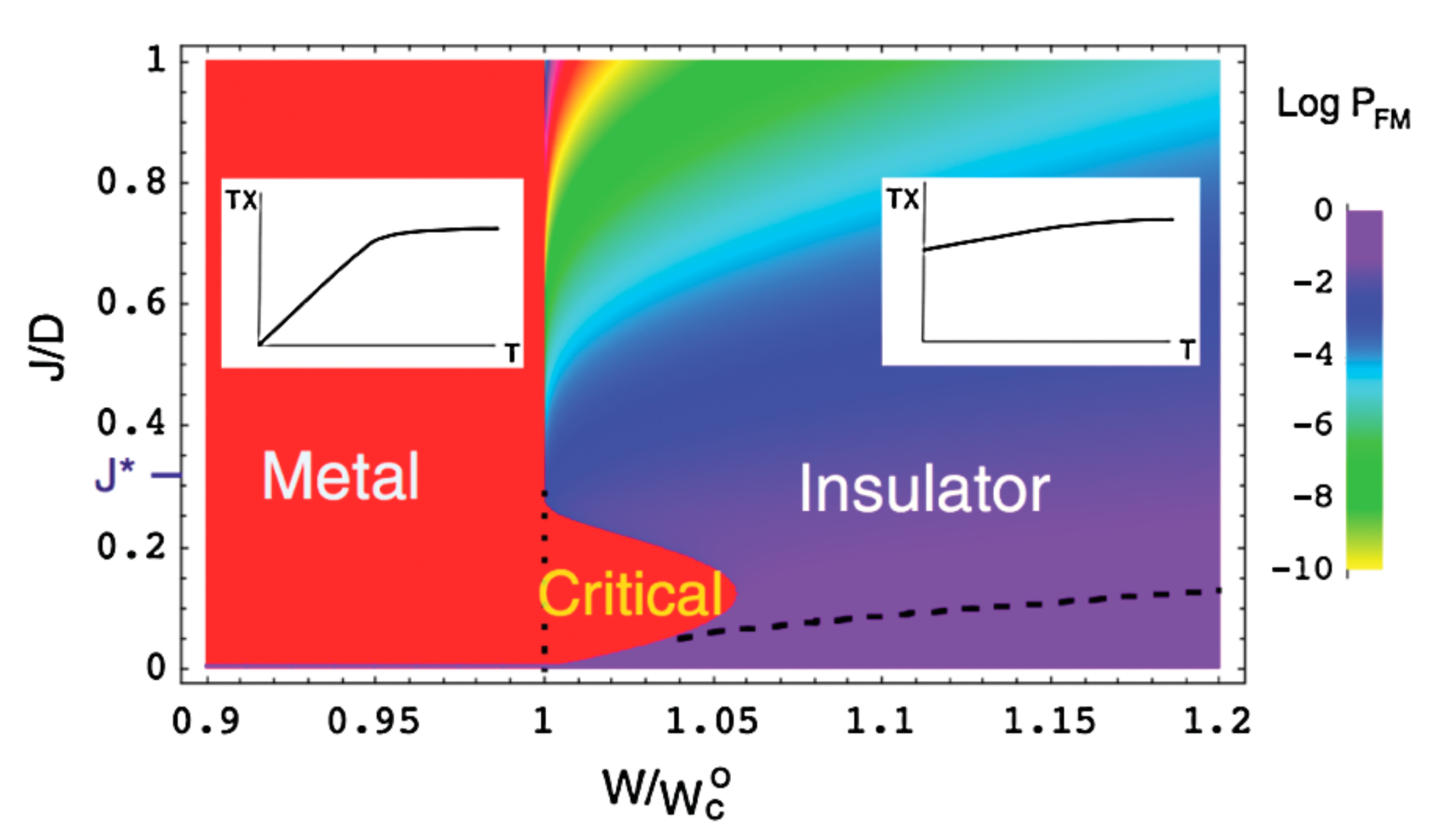}
\caption{ 
 Fraction of free MMs $P_{\rm
    FM}$ at $T=0K$ in a 3D disordered metal as
  function of  exchange coupling $J$ (in units of  band width
  $D_0$) and disorder strength $W$ (in units of  orthogonal critical value
  $W_{\rm c}^{\rm O} $). Critical correlations result in  finite
  $P_{\rm FM}$ even for $J > J_c^A$, Eq. (\ref{jca}) (dashed line). For $J <
J^\ast $ ( as given by Eq. (\ref{jstar})) there is a critical region for disorder
amplitudes $W_{\rm c}^{\rm O} < W < W_{\rm c}^I (J),$ which is  
given by Eq. (\ref{wcj}). 
Insets: schematic $T-$dependence of magnetic suseptibility $\chi$  times $T$. 
Fig. taken from   Ref.  \cite{PRL09},  Copyright  2009, American Physical Society. }
\label{kats}
\end{center}
\end{figure} 

 At finite temperature $T$  and in the presence of MMs     the  scaling Ansatz 
 for   the  conductivity  becomes therefore 
      \begin{equation} \label{scaling2}
       \sigma(n,T) =
 \Delta n^{(d-2)\nu}  F [ T  \Delta n^{-z \nu},\tau_s^{-1}(T) \, \Delta n^{-\varphi_s}].
 \end{equation}
    Since MMs are Kondo screened in the metallic phase, 
     while the Kondo screening can become quenched in the localised phase
     by the localisation length $\xi_c,$
      the  spin scattering rate can be  different in the metallic phase than in the insulator where
       free MMs remain.
      Therefore, in Refs.  \cite{PRL09,Kats} it was pointed out that 
       there can emerge a new critical phase. The resulting quantum phase diagram was derived
        analytically, as shown in  
     Fig.  \ref{kats}, where the concentration of free  MMs at zero temperature is plotted.
    Furthermore, the temperature dependence of the spin relaxation rate results in   finite temperature delocalization transitions \cite{PRL09,Kats}.
 In Ref.  \cite{Kats} in the limit of  dilute MMs 
the shift of the
critical disorder as function of exchange coupling $j=J/D$ was found to be given by 
\begin{equation}
\label{wcj}
W^I_{\rm c} (j)= W_{\rm c}^{\rm O} + W_{\rm c}^{\rm O} \left(
\frac{a_{c}^2}{D_{e} \tau^{0}_{s}}\right)^{\kappa(j)} ,
\end{equation}
where $1/\kappa(j) =\varphi_s - 2 \nu \frac{d^2}{2   \eta }( j +  \eta/d )^2$. 
Here,  the deviation of scaling 
 from $\varphi_s$ arises from the fact that the density of free MMs depends on 
localisation length $\xi \sim (W-W_c)^{-\nu},$ itself.
This result is valid for dilute magnetic impurities, dominated by the Kondo effect and for small deviations from the orthogonal class value $W_{\rm c}^{\rm O}$.
 For larger exchange couplings  $j > j^\ast$, where
  \begin{equation} \label{jstar}
  j^\ast =
 (2 \sqrt{  \eta}-  \eta)/d,
\end{equation}
 (which is in $d=3,$    $j^\ast \approx 0.276$), 
  the Kondo screening is no longer quenched, 
 and  the critical disorder  $W^I_{\rm c}$ approaches  the one of the orthogonal symmetry class $W_{\rm
  c}^{\rm O}.$

   For a
clean metallic system of finite size with finite number of states $N$, 
  the Kondo renormalization  is cutoff by the mean level spacing $\Delta =D_0/N$.
   The critical exchange coupling $J_c$ below which 
 the Kondo screening is quenched
 and the MMs remain free, 
vanishes then logarithmically with the number of states in the band, $N,$ as $J_c \sim
D_0/\ln N$. In an Anderson insulator  the eigenstates at the Fermi energy are localized with a
localization length $\xi.$ Thus,  there are  finite local gaps of order $\Delta_{\xi_c} =
(\rho\, \xi^d)^{-1}$  at the Fermi energy which cut off the Kondo
renormalization  and there are  free MMs whenever
$J < J_c^{\rm A}$, where 
\begin{equation} \label{jca}
J_c^{\rm A}  =  D_0/\ln
N_{\xi_c},
\end{equation}
 with $N_{\xi_c}= D_0/\Delta_{\xi_c}$  the number of localized
states with a finite wave function amplitude at the site of the MM  \cite{Zhuravlev2007}.
 However, Eq. (\ref{jca}) does not take into
account multifractality and  critical correlations between wave
functions at different energies at the AMIT  \cite{powerlaw}. In fact, the 
amplitude of multifractal states can be  suppressed at some random
positions below their typical value, scaling as $L^{-\alpha_{\psi}}$ with
$\alpha_{\psi} > d$ (i.e., decaying faster with system size $L$ than extended
states). Correlations between wave functions at different energies 
then open wide local pseudogaps \cite{Zhuravlev2007,Kats}. 
  The wave function intensities within a
localization volume is close to 
 log-normal distribution  with $\alpha_{\psi} \rightarrow \alpha_{\psi,\xi} =
- \ln |\psi|^2/\ln \xi$\cite{Kats}. For the evaluation of
$J_c$  the system length $L$ needs then to be 
substituted by the localization length $\xi(W) \sim
(W- W_c)^{- \nu}$. Thus, for fixed $J$, the density of free MMs 
 are found to depend on the localisation length $\xi$ as \cite{Kats}
\begin{equation}
\label{eq:nFM}
P_{\rm FM} = n_{\rm FM} /n_M = {\rm Erfc} \left( \sqrt{\frac{\ln
\xi}{2 \eta/d}}  \frac{J}{D} \right).
\end{equation}
Close to the transition,  where $\xi$ is large,  the density of free MMs 
Eq. (\ref{eq:nFM}) relative to $n_M$ simplifies to  $P_{\rm FM} \sim (W -W_c)^{
\kappa(J)}$ with  positive exponent $\kappa(J) = (\nu /2 \eta) (
J/D)^2$. $P_{\rm FM}$ is plotted in Fig. \ref{kats}
as function of disorder strength $W$ and exchange coupling $J$. 
  It is vanishing both in the  metal and critical regions due to Kondo screening.
  In 
  the insulator region it remains finite due to the quenching of the 
   Kondo effect by localization and local pseudogaps. 
   The critical region appears because for 
 $j<j^\ast$  the position of the critical point $W_{\rm c}$ depends on the
direction from which the AMIT is approached. Thus there exists a {\it critical phase} for 
intermediate disorder strengths $W_{\rm c}^{\rm O}< W < W^I_{\rm c}
(j)$ .   The width of that 
critical phase is $W^I_{\rm c} (J) - W_{\rm c}^{\rm O} \sim 
 n_{M}^{\kappa(j)},$ increasing with a power of the density of MMs, $n_{M}$ \cite{Kats}.
 
 {\it Magnetic field. }
 One way to probe that quantum phase diagram is, to apply 
  a  magnetic field which  polarizes the free MMs
 reducing thereby  the 
   spin relaxation rate  \cite{bobkov,vavilov,micklitz07}.
 Also, the  Kondo singlet   is  partially broken up by  the Zeeman  field. Thus,
 MMs
     contribute a
      spin relaxation rate which increases 
       with the Zeeman field.  
     An orbital magnetic field  breaks  time reversal symmetry
        and therefore  results also in a   shift of the AMIT,  approaching  the unitary limit.   
We found in Ref.  \cite{Kats} that the {\it Zeeman field }
    shifts  the critical disorder to
    \begin{equation} \label{wcb} 
    W_{c}(B) = W_{c}^I(j) + W_{c}^O c_{M} \left(\frac{  \gamma _{s}  B |S_{z}|}{E_{c}}\right)^{j \kappa(j)},
    \end{equation}
 where
  $c_{M} = \left(d j n_{M}/((2-j) \pi \rho D_{e})\right)^{ \kappa(j)}$.  
      Thus, one finds that the transition between the critical  phase  and the insulator  is shifted 
       in a magnetic field according to Eq. (\ref{wcb}).
   The {\it orbital  magnetic field} is known \cite{Khmelnitskii1981} to shift $W_{c}$ to 
    \begin{equation} 
    W_{c}(B)  = W_{c}^O + W_{c}^O ( \pi e B/h)^{1/(2\nu)}.
    \end{equation}
     This determines the transition line between metal and critical phase, 
      since the Zeeman field   contributes a slower dependence on $B$, 
     coming from the metal side of the transition.
  For the transition between critical phase  and insulator, we  find  that for 
$
  j < j_{Z} =   \eta/d  \left(  2 \sqrt{d+1 +d^2/\eta }/d -1-2/d \right),
$
   the shift of $W_{c}$ is dominated by the Zeeman field. 
  For, $ d=3,   \eta/d  =2/3$ one finds, $j_{Z}= 0.185$. Thus the  Zeeman field 
   is dominating the shift of the transition
    for  realistic values of exchange couplings $j$.

\section{ Finite temperature Delocalization Transitions} Since  the spin relaxation rate   depends on
temperature,  Eq. (\ref{eq:tauapprox}),  the  breaking of the 
 spin- and time-reversal-symmetry is found to  depend on
temperature, as well. Since we have seen above that 
the position of the AMIT is determined by  the spin relaxation rate, it 
shifts as function of temperature $T$. Therefore, the metal-insulator
transition  can occur at a finite temperature $T_{c}(W,J)$. In order to
 investigate the existence of such a transition, we apply  the
Larkin-Khmel'nitskii condition as outlined above \cite{Khmelnitskii1981,Droese1998}  with the temperature
dependent  symmetry parameter
$X_{s} (T) = \xi^2/D_{e} \tau_{s}(T)$ \cite{Kats}.

{\it  Approaching the AMIT from the Insulator side.}
Coming from the insulating side of the transition, where   the
localization length $\xi$ is still finite and 
 smaller than the thermal length $L_{T}$, 
  the ratio
  $X_{s} (T)$ is finite. $1/\tau_{s}(T)$ saturates
at low temperatures to the spin relaxation rate from free MMs.  Thus, at low temperatures
 the transition occurs at $W_{c}^I(J),$ as given by Eq. (\ref{wcj}).
 Since the Kondo temperature is distributed, the  spin relaxation rate at finite temperature 
    is given by a weighted integral over its distribution function  $P(T_K)$
 as $1/\tau_{s} (T) = \int_{0}^{\infty} d T_{K} P(T_{K}) 1/\tau_{s} (T/T_{K}) $.
  The spin relaxation rate of a magnetic impurity with a given Kondo temperature
   $T_{K}$, $ 1/\tau_{s} (T/T_{K})$,
 is given by Eq. (\ref{eq:tauapprox}). Thus, it 
    increases first like $T^2/T_{K}^2$ 
    when $T < T_{K}$, 
reaching  a maximum at   $T_{K}$  before it decays logarithmically towards its
saturation value $1/\tau^{0}_{s}$.  For  low temperatures $T \ll T^0_{K}$, one  can 
  approximate
$1/\tau_{s} (T)$ by a sum of the spin relaxation rate from 
  free MMs  at sites whose density of states is suppressed as $\rho({\bf r})\sim \xi^{d-\alpha_{FM}}$,
   and the spin relaxation from spins whose Kondo temperature exceeds $T$.   
\begin{figure}[t]
\includegraphics[width=8.2cm]{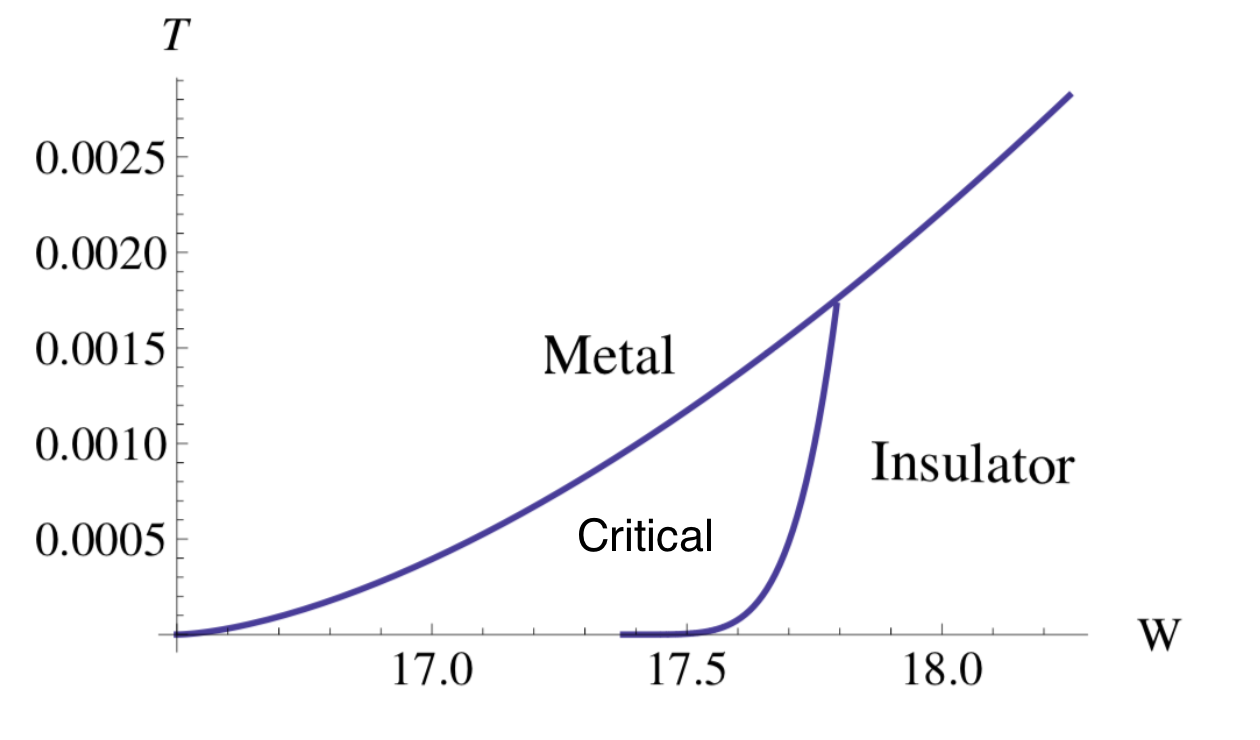}
\caption{(Color online) Finite-temperature phase diagram of
  Kondo-Anderson transitions. Plots of the
  critical temperatures $T_{c}^M(W,J) $, Eq. (\ref{tcm}) (upper solid line), and
  $T_{c}^I(W,J)$, Eq. (\ref{tci}) (lower solid line), respectively, with disorder
  amplitude $W$  in units of hopping parameter $t$ and 
  temperature  in units of   $E_{c}$, using 
  $j =0.2$, $\alpha_{0}=4$, $d=3$, $  \eta/d  =2/3$, $\nu=1.57$ and
  $a_{c}^2/(D_{e}\tau^{(0)}_{s}) = 0.1$.   Fig. taken  from 
 Ref.   \cite{Kats},  Copyright  2012, American Physical Society. }
\label{fig:Tphasediagram}
\end{figure}
 \begin{figure}
\begin{center}
\includegraphics[width=0.45\textwidth]{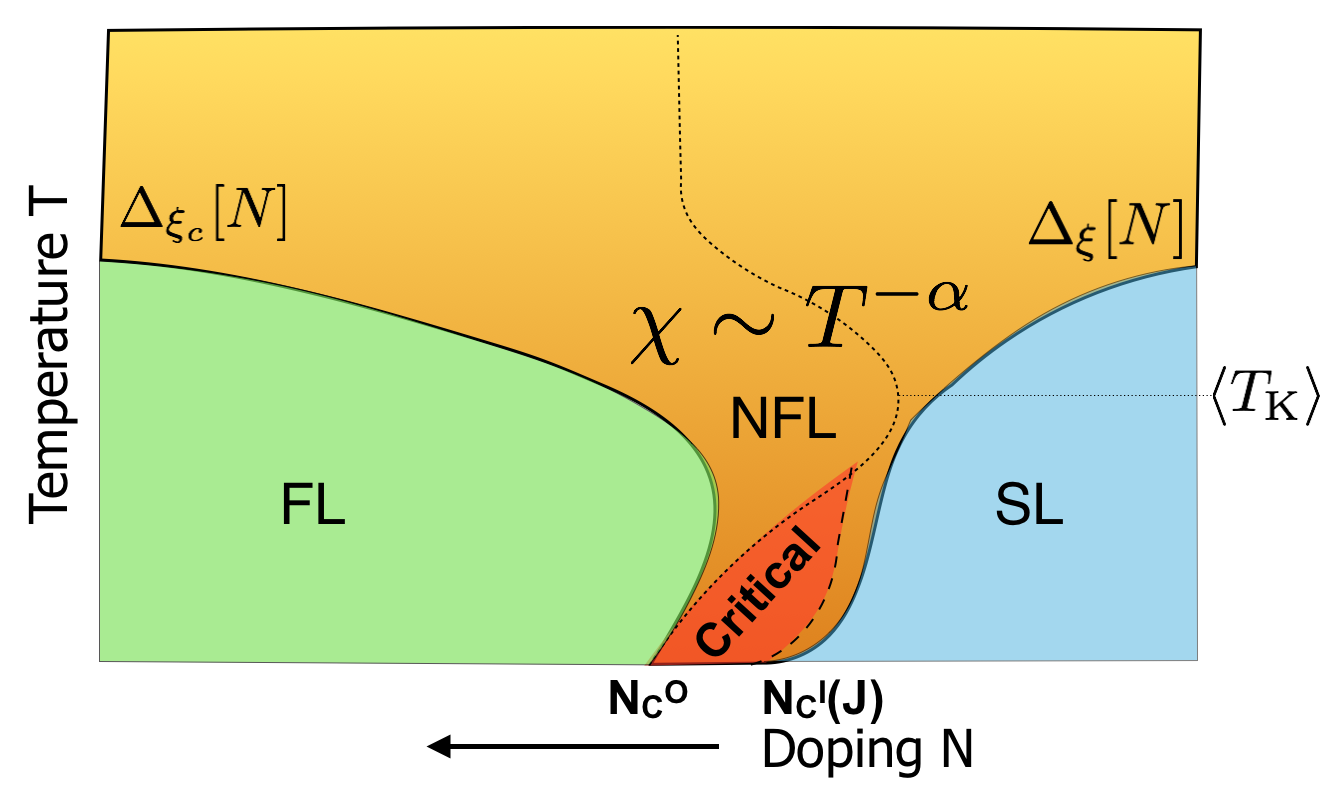}
\caption{ 
 Schematic phase diagram,  with the  crossover from Fermi liquid (FL) 
 to non-FL behavior and a critical phase.
   The dashed lines
 are sketches of  the critical concentration $N_c(T)$ which depends on 
 temperature due  to the temperature dependent spin scattering rate. 
 It has a maximum at the average Kondo temperature  $\langle T_{K} \rangle,$ which is indicated by the dotted line. 
 The critical phase forms since the spin scattering is different when coming from the metal, where MMs are Kondo screened  and  from the insulator side of the transition, where there are free MMs.  }
\label{kats2}
\end{center}
\end{figure} 
From the scaling with $X_s(T)$ we can derive, as outlined above, 
 the critical temperature $W_{c}^I(J,T)$, which thereby depends on temperature $T$.
 As a consequence,  we find a finite 
  transition temperature between the insulator and an extended phase as 
\begin{equation} \label{tci}
T_{c}^I =E_{c } c_{I}\left(
\frac{W-W_{c}^{I}}{W_{c}^{O}} \right)^{\frac{1}{ j
      }},
\end{equation}
 where
 $W_{c}^{I}$ is given by Eq. (\ref{wcj}), and 
  $c_{I}=  (\kappa_{j}/j)^{-1/j} (D_{e} \tau_{s}^0)^{\kappa_{j}/j}$,
   where $1/\kappa_{j} = \nu (2-d^2/2/\eta (j+\eta/d)^2)$.
   Eq.  (\ref{tci}) is plotted in Fig. \ref{fig:Tphasediagram} (lower solid line).

{\it Approaching the AMIT from the Metallic side.}
Coming from the
metallic side, the density of free MMs  decays
 at  low temperatures $T < \Delta_{\xi}$ to zero.
  There,  the spin relaxation rate is dominated by partially screened  MMs
  with $T < T_{K}$. Thus,  
the transition is shifted to a different value, 
$W_{c}^M (T).$ Accordingly we  find  the   transition temperature to the metal as 
\begin{equation} \label{tcm}
T_{c}^M (W) = \sqrt{ S(S+1) \pi^2 D_{e} \tau_{s}^0 } 
\left| \frac{W-W_{c}^{O} }{ W_{c}^{O} } \right|^{  \nu } T_{K}^{0},
\end{equation}
as  is plotted in Fig. \ref{fig:Tphasediagram} (upper solid line) as function of disorder amplitude $W$.

Thus, we can conclude that  in the dilute MM limit 
the temperature dependence of the spin scattering rate results in 
 finite temperature  transitions with  critical temperatures 
 $T_{c}^M(W),$ 
  Eq.   (\ref{tcm}) and $T_{c}^I(W)$,
   Eq.  (\ref{tci}).
   Since $T_{c}^M(W) \neq T_{c}^I(W)$ there is 
a critical  region, as seen in  Fig. \ref{fig:Tphasediagram}.
 The corresponding finite temperature phase diagram as function of doping concentration $N$ is shown in
 Fig. \ref{kats2}, schematically. 
 
 We note that this phase diagram  is further complicated by
the fact that $1/\tau_{s}$ reaches a maximum and decays to its saturation value
logarithmically at temperatures exceeding the average Kondo temperature  $\langle T_{K} \rangle$,
as indicated schematically 
 by the dotted line in Fig. \ref{kats2}. Furthermore, at higher temperature  the phase
boundaries are  less well defined  because of  
 inelastic scattering and dephasing processes.
  \begin{figure}
\begin{center}
\includegraphics[width=0.45\textwidth]{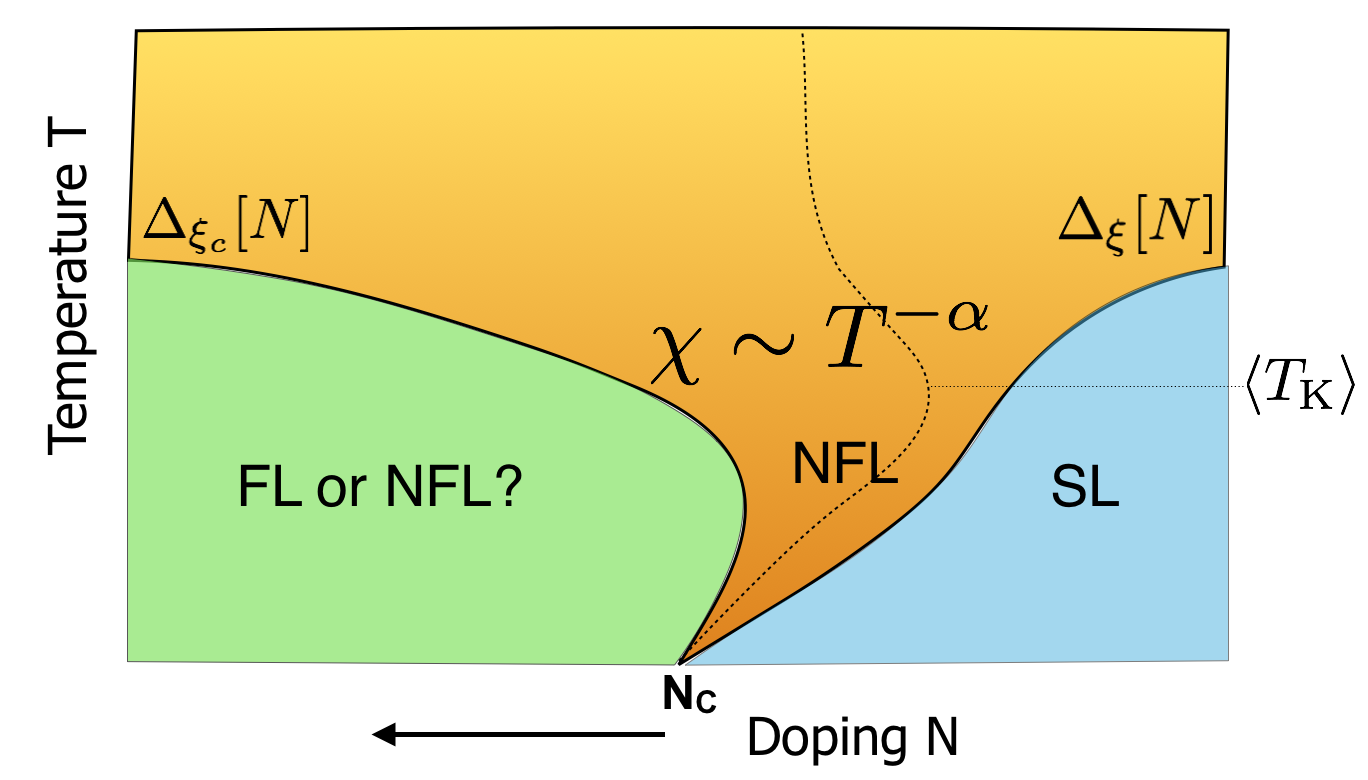}
\caption{ 
 Schematic phase diagram, in the presence of a finite concentration of MMs
 as function of doping concentration $N$ and temperature $T$.
  The dashed line indicates the temperature dependence of the critical doping concentration 
  $N_c(T)$ due to the temperature dependence of the spin scattering rate,    $1/\tau_s(T)$
  (It has a maximum at the average Kondo temperature  $\langle T_{K} \rangle,$ which is indicated by the dotted line.) This 
  results in finite temperature transitions between the localised and  the non-Fermi-liquid 
  metallic phase. In the metallic phase, the  competition between Kondo screening of local moments and indirect exchange coupling may in the low temperature limit $T  < \Delta_{\xi_c}$ either lead to a disordered Fermi liquid or non-Fermi liquid, as indicated by  the green area, in addition to the Althshuler-Aronov type corrections due to long range interactions\cite{altshuleraronov,finkelstein,finkelstein88}.}
\label{true}
\end{center}
\end{figure} 

{\it Finite Temperature Phase Diagram with Coupled MMs. } The  phase diagram shown in  Fig. \ref{kats2} is expected to be  modified 
    when 
      the indirect exchange  couplings between MMs dominate the Kondo effect for a finite 
      concentration of MMs. 
        Implementing  both the  Kondo screening and   the effect of the 
      RKKY couplings on the temperature dependence of the 
       spin relaxation rate $1/\tau_s (T),$ as well as   the doping dependence
     of the concentration of MM's, $n_M(N),$  as reviewed above, 
     the finite temperature phase diagram
     of doped semiconductors can be derived. 
     Furthermore, as reviewed above,   
         there are strong indications that for any finite number of ferromagnetic  couplings, as they 
         are pressent close to the MIT, 
       the coupled quantum spin system is driven to a fixed point with  clusters forming  large effective spins, contributing  a  Curie law magnetic susceptibility \cite{sigrist2,Wan02}.  
       Thus, there might be 
       a finite concentration of such large effective spin clusters  on both sides of the transition, which make the spin scattering rate $1/\tau_s$ finite. 
        Then,  the critical tongue found in the dilute MM limit as shown in 
        Fig.  \ref{kats2}, will be  diminished at finite density of MMs. 
     Still, as there will be a  coexistence with a  finite density of 
     Kondo singlets and of random singlets, as reviewed above, the 
      spin scattering rate 
     $1/\tau_s(T)$ is expected to increase with temperature 
     since the  Kondo screened MMs and the random 
  singlets, which do not contribute to spin scattering at low temperature,
  are increasingly   broken up with increasing   temperature $T$. Fig. \ref{true}
   shows a sketch of the resulting expected  phase diagram as function of doping concentration $N$ 
   and temperature $T$. 
    Furthermore, the presence of clusters forming  large effective spins,  which contribute
      a  Curie law magnetic susceptibility \cite{sigrist2,Wan02}, and possibly also  the presence of free MMs due to 
     rare sites   of dopant sites,  which remain strongly isolated on the metallic side of the transition
      \cite{BhattFisher92}, 
     may  result in a divergence of the magentic susceptibility in the low temperature limit and non-Fermi liquid behavior, even at low temperatures $T < \Delta_{\xi_c}$, as indicated 
     by a question mark in Fig. \ref{true},  in addition to the Althshuler-Aronov type corrections due to long range interactions\cite{altshuleraronov,finkelstein,finkelstein88}.
     This  depends on the type of doping, and whether dopant locations are uncorrelated, as assumed in Ref.   \cite{BhattFisher92}  or so much correlated, 
     that well isolated sites do not occur on the metallic side of the MIT.

\section{Conclusions and Outlook}
     
   While  still more experimental and theoretical 
     research on doped semiconductors in the vicinity of the MIT is needed,  
    we can conclude that 
     it  is now very well established that a finite density of 
     magnetic moments are present in the vicinity of the MIT which need to be taken into account in a comprehensive theory of their MIT. 
      These  randomly positioned MMs  can couple to form a spin liquid phase.
       This spin liquid phase is well established to occur in the insulating phase 
       at low doping below the MIT, and 
        at low temperatures, as
   shown schematically  in blue in the 
   phase diagram Fig. \ref{true}. 
      This spin liquid  is   described very well by the 
 Bhatt and Lee  theory, where  clusters of randomly coupled spins form,
 which, due to the random antiferromagnetic interactions at low doping, 
 are  mostly random singlets\cite{bhattlee81,bhattlee82}. The excitations of these 
  random clusters of spins at finite temperature  
result in  the anomalous power law divergence of   the 
  magnetic susceptibility   $\chi (T) \sim T^{-\alpha},$ with  $\alpha \le 1$ \cite{bhattlee81,bhattlee82}.

     There is furthermore evidence for  another region,  the fan colored  in orange in the 
     phase diagram Fig. \ref{true}, where the  doped semiconductor have also 
     non-Fermi-liquid properties, 
     as characterised, also by a  divergent magnetic susceptibility, but 
     which may be  dominated by the formation of randomly distributed Kondo singlets. 
  The  agreement of the  anomalous power  $\alpha$ with the universal value Eq. (\ref{universal}), 
  $\alpha= 1- (\alpha_0-3)/3$, independent on the doping concentration $N$ within this fan region, 
   with  the experimental results,  as shown in the  inset (circles)  of 
   Fig. \ref{suscsip}  \cite{Schlager1997} and  results reported in Ref.     \cite{bhatt88},
   is striking.
This universal value has been 
derived 
  by taking into account  multifractal correlations in the calculation of the local Kondo temperatures,  which yields a   distribution of Kondo temperatures  with  a power law tail 
   Eq.  (\ref{universal}) both at the MIT and in its vicinity on both sides of the transition. 
      The fact that the resulting  anomalous power of the magnetic susceptibility 
     is experimentally  found to be  independent of doping close to the MIT  and 
   in good agreement with    Eq.  (\ref{universal}), as derived from multifractality, 
   might make these experiments 
    the first, albeit indirect, measurements of multifractality at the MIT in doped semiconductors. 
  
  Furthermore,  temperature dependent spin correlations are found to cause 
        finite temperature transitions between a localised and   non-Fermi-liquid 
  metallic phase. The critical doping concentration is found to depend on temperature, 
  $N_c(T),$ due to  the temperature dependence of the spin scattering rate    $1/\tau_s(T),$ 
   as indicated in Fig. \ref{true}, dotted line, which is caused by temperature dependent  spin correlations, such as the Kondo effect.
    The question arises how to detect these finite temperature
    MITs experimentally. Neglecting the coupling to phonons, one would have a
     transition between zero conductivity $\sigma(T) =0$ at $T < T_c(N,J)$ and 
     a metal with finite conductivity  $\sigma(T) >0$ at $T > T_c(N,J).$
      Thus, that makes  it similar to the 
      finite temperature many body localization (MBL) transitions, which were
       suggested in Refs  \cite{basko} and 
       \cite{gornyi} and implied in the  work Ref.   \cite{Fleishman1980}, where arguments were presented that short-range interactions in an electron system with localized single-particle states might not destroy localization for some range of finite temperature $T$.
      MBL is  meanwhile also understood to manifest itself by the lack of thermalization, see  Refs.   \cite{Ogan2007,Pal2010}. 
      With phonon scattering one expects for   $T < T_c(N,J)$
    an exponentially low conductivity, as described by   the  Efros-Shklovskii  variable range hopping 
 conductivity \cite{efros}. 
  The application of a magnetic field is expected to 
  strongly change the phase diagram and thereby lead to giant magneto conductivity, 
   since  the critical disorder,  Eq. (\ref{wcb}), and respectively, the critical dopant density of 
    the transition depends on the magnetic field. 
    Further theoretical research and experiments on doped semiconductors will be needed
     to explore and understand 
     the finite temperature diagram as function of doping and magnetic field completely
     and resolve remaining  
  open problems which  include: 
      It is not yet clear whether and   under which conditions  the  
          random Kondo singlet phase can become quenched by indirect exchange interactions
           between magnetic moments. 
          While there is strong evidence that 
             the supercritical low temperature phase is   dominated by 
       long range Coulomb interactions which are only partially screened due to disorder \cite{Schwiete2023},  it remains to be understood under which conditions 
       a spin liquid phase  of  magnetic moments may persist in this disordered metal phase\cite{BhattFisher92}.

\section*{Acknowledgments}
  This review reflects my personal  perspective achieved through fruitful  
 collaborations and discussions with Ravin Bhatt, Georges Bouzerar, Konstantin B. Efetov, Peter Fulde, 
 Jean-Francois Guillemoles, 
 Stephan Haas, Ki-Seok Kim,  Bernhard Kramer, Alexander Lichtenstein, Hu-Jong Lee, 
 Hyun-Yong Lee, Youcef Mohdeb, 
 Eduardo R. Mucciolo, Tomi Ohtsuki, Mikhail Raikh, Vincent Sacksteder, Keith Slevin, Alexey Tsvelick, Javad Vahedi, Imre Varga and  Isa Zharekeshev, which I gratefully acknowledge, 
and  thanks to insightful  discussions with Elihu Abrahams, Kamran Behnia,  Claudio Chamon, 
Vladimir Dobrosavljevic, Alexander M. Finkel'shtein, Serge Florens,  Hans Kroha, Raj Mohanty, C\'ecile Monthus,  Philippe Nozi\`eres and Christoph Strunk. 
I am grateful to Ki-Seok Kim and  Tomi Ohtsuki for critical reading of the manuscript and useful comments. 
I am particularly grateful to Konstantin B. Efetov for introducing me to the fascinating field of disordered systems while being his PhD student and then his Post Doc, and for his  guidance and  comprehensive advice \cite{advice} ever since. 
 Support from DFG (Deutsche Forschungsgemeinschaft) KE-807/22-1 is gratefully acknowledged.

\end{document}